\documentclass[reprint, amsmath,amssymb,aps,onecolumn]{revtex4-2}

\usepackage{graphicx}
\usepackage{bm, slashed}
\usepackage{tikz} 
\usepackage{hyperref} 
\usepackage{comment}
\usepackage{mathdots,bbold}
\usepackage{soul}
\usepackage[normalem]{ulem}
\setcitestyle{square,numbers}
\usepackage{pgfplotstable}

\begin{document}

\title{Critical magnetic flux for Weyl points in the three-dimensional Hofstadter model}

\author{Pierpaolo Fontana}
 \email{Pierpaolo.Fontana@uab.cat}
\affiliation{
 Departament de Física, Universitat Aut\`{o}noma de Barcelona, 08193 Bellaterra, Spain
}

\author{Andrea Trombettoni}
%\email{atrombettoni@units.it}
\affiliation{%
 Department of Physics, University of Trieste, Strada Costiera 11, I-34151 Trieste, Italy
}
\affiliation{SISSA and INFN, Sezione di Trieste, Via Bonomea 265,
I-34136 Trieste, Italy}

\date{\today}

\begin{abstract}
    We investigate the band structure of the three-dimensional Hofstadter model on cubic lattices, with an isotropic magnetic field oriented along the diagonal of the cube with flux $\Phi=2 \pi \cdot m /n$, where $m,n$ are co-prime integers. Using reduced exact diagonalization in momentum space, we show that, at fixed $m$, there exists an integer $n(m)$ associated with a specific value of the magnetic flux, that we denote by $\Phi_c(m) \equiv 2 \pi \cdot m/n(m)$, separating two different regimes. The first one, for fluxes $\Phi<\Phi_c(m)$, is characterized by complete band overlaps, while the second one, for $\Phi>\Phi_c(m)$, features isolated band touching points in the density of states and Weyl points between the $m$- and the $(m+1)$-th bands. In the Hasegawa gauge, the minimum of the $(m+1)$-th band abruptly moves at the critical flux $\Phi_c(m)$ from $k_z=0$ to $k_z=\pi$. We then argue that the limit for large $m$ of $\Phi_c(m)$ exists and it is finite: $\lim_{m\to \infty} \Phi_c(m) \equiv \Phi_c$. Our estimate is $\Phi_c/2\pi=0.1296(1)$. Based on the values of $n(m)$ determined for integers $m\leq60$, we propose a mathematical conjecture for the form of $\Phi_c(m)$ to be used in the large-$m$ limit. The asymptotic critical flux obtained using this conjecture is $\Phi_c^{{\rm (conj)}}/2\pi=7/54$.
\end{abstract}

\maketitle

\section{\label{intro}Introduction}
% Broad introduction
The interaction between charged particles and background magnetic fields plays an important role in the realms of condensed matter physics and solid-state theory \cite{Landau1981,Grosso_Parravicini_2000,El-Batanouny2020}. The associated phenomena have been largely studied in the last decades, leading to the discovery of mathematical models that have changed our understanding of electronic properties in crystalline materials. In this context, one of the most fascinating and paradigmatic models is the Hofstadter model \cite{Azbel1964,HofstadterPRB1976}, which combines the interaction of electrons in the lattice structure of solids with external magnetic fields. During the years, this model has become increasingly important for its complex and rich structure, both from the physical and mathematical points of view.

% Importance in physics: 2d
In two dimensions, the tight-binding model combined with the Peierls substitution \cite{Peierls1933} captures how the properties of electrons moving in a periodic lattice are influenced and modified by the presence of a perpendicular magnetic field. The first studies were focused on the broadening of energy levels and the determination of the wave function of the system using semi-classical techniques, reducing the problem to a finite difference equation known as the Harper equation \cite{HarperPPS1955,Sokoloff1985,Thouless1990,KohmotoPRB1994}. The complete band structure as a function of the magnetic flux has an intricate self-similar structure, which is the celebrated fractal Hofstadter butterfly \cite{HofstadterPRB1976}. Moreover, the two-dimensional (2D) Hofstadter model is closely connected to the physics of Chern insulators and quantum Hall effect \cite{vonKlitzingPRL1980,Yoshioka2002,Harper&RoyPRB2014}, due to the possibility of realizing novel topological phases of matter by introducing strong magnetic fluxes in 2D materials \cite{SantosPRL2020,BernevigPRL2020}.

% Importance in physics: 3d and beyond
Interesting features emerge as well in the three-dimensional (3D) case, showing a highly complex band structure depending on the hopping strengths in the various directions. For strong anisotropic hoppings along a particular spatial axis there exists a 3D analog of the Hofstadter butterfly \cite{KoshinoPRL2001}, with a fractal structure of the energy levels as a function of the magnetic field tilt with respect to the anisotropy axis. For general hopping coefficients or orientations of the magnetic field, however, this structure disappears and the spectrum is no more gapped. The general absence of energy gaps between the bands is the reason why the connection with the quantum Hall physics can not be trivially extended to the 3D case \cite{KunsztZeePRB1991,KoshinoPRB2003}. The 3D Hofstadter model has been studied for varying directions and 
intensities of the magnetic flux \cite{Hasegawa1990,LaughlinZouPRB1990,KunsztZeePRB1991,Hasegawa1992,KoshinoPRL2001,KoshinoPRB2003,FontanaPRB2021}. Analogous to the behaviour observed in the 2D Hofstadter model in the $\pi$-flux case \cite{Affleck-MarstonPRB1989}, the 3D case also exhibits points in the spectrum with zero energy density of states and band touching, such as those found with a $\pi$-flux \cite{LaughlinZouPRB1990} or for fluxes of the form $\Phi=2\pi / n$ with $n$ integer  \cite{Hasegawa1990,Hasegawa1992,FontanaPRB2021}. Due to these properties, the 3D Hofstadter model can be used to model various topological metallic phases, such as Weyl metals and semimetals \cite{Wan-Vinshwanath-Savrasov(PRB),Burkov2011,BurkovReview2018}. Interestingly, the semimetal behaviour persists even at finite interaction strengths, below a critical value \cite{Sorella_1992,Mazzucchi_2013}, and in presence of anisotropic tunneling coefficients \cite{Hasegawa1990,Mazzucchi_2013}. In addition to the 3D case, higher dimensional extensions of the Hofstadter model have been considered in relation to the investigation of integer quantum Hall effect in even dimensions \cite{Kimura2014}.

% Experimental realizations
The features of the Hofstadter model can be explored experimentally in various platforms, also due to the recent remarkable progress in the engineering of quantum matter in artificial lattices in presence of large magnetic fluxes \cite{Aidelsburger2018}. Recent studies includes photonic crystals \cite{ozawa2019}, ultracold atoms in optical lattices \cite{Aidelsburger2013,Miyake2013,goldman2014,Aidelsburger2015,Weitenberg2021}, moiré double-layter heterostructures and superlattices \cite{Dean2013,Hunt2013}, and molecular nanostructures built with scanning tunneling microscopes \cite{kempkes2019,fremling2020}. Moreover, the topological properties of the Hofstadter model in different dimensionalities have interesting connections with optical diffraction figures resulting from optical gratings \cite{DiColandrea2022}.

% Importance in mathematics
Besides these physical features and applications to the analysis of new phases of matter, the Hofstadter model displays a rich mathematical structure interesting \textit{per se}, intrinsically connected to incommensurability effects \cite{HofstadterPRB1976,Obermair1976,Wannier1978,Aubry1980,Guillement1989} and topological invariants \cite{ThoulessPRL1982}. The Harper equation represents a particular type of almost Mathieu operator, and its properties and solutions strongly depends on the rational nature of the magnetic flux. Indeed, if the flux is an irrational number, it can be proven that the spectrum of the model is a Cantor set \cite{Avila2009,Bruning2007}. Conversely, for rational fluxes the spectrum has a finite number of bands and can be investigated in the reciprocal space \cite{KohmotoPRB1989,WenNPB1989}, by taking advantage of the Bloch theorem applied to the magnetic translation group \cite{ZakIPRA1964,ZakIIPRA1964,Rauh1974}.

% Our setting: 3D and rational flux
Throughout this paper, we focus on the 3D case and rational magnetic fluxes parametrized as 
\begin{equation}
    \Phi= 2 \pi \cdot \frac{m}{n}\,,
    \label{phi}
\end{equation}
where $m,\;n\in\mathbb{N}$ are co-prime integers. The spectrum consists of the union of $n$ bands. By excluding the case of strong anisotropic hopping, previous works with this setting suggest that these bands touch in isolated points or overlap, depending on the value of the magnetic flux \cite{Hasegawa1990,Hasegawa1992,FontanaPRB2021}. Investigations for small values of $n$ and $m$ indicate that for large fluxes the bands touch, and by progressively decreasing $\Phi$ they overlap 
\cite{Hasegawa1990,Hasegawa1992}. We may then define, at fixed $m$, the \textit{critical flux} $\Phi_c$(m) as the magnetic flux at which this transition from isolated touching to full bands overlap possibly takes place. For example, for $m=1$, Weyl points appear if $n=7$, while bands overlap occurs for $n=8$ \cite{Hasegawa1990,FontanaPRB2021}. Similarly, for $m=4$ a transition occurs between $n=31$ and $n=32$ \cite{Hasegawa1990}. Moreover, for $m=1$, the Weyl points are observed between the first and second bands \cite{Hasegawa1990,LaughlinZouPRB1990,FontanaPRB2021}. It is expected that for fluxes $\Phi>\Phi_c(m)$, Weyl points exists between the $m$- and the $(m+1)$-th band. 

The questions we address in the present paper are the following ones: {\it i)} is the critical flux $\Phi_c(m)$  defined for any $m$? {\it ii)} If the the answer to the previous question is affirmative, does the limit for large $m$ converge to a well-defined finite flux $\Phi_c$? Additionally, we aim to investigate whether, as expected, the Weyl points separate the $m$- and the $(m+1)$-th bands, for $\Phi>\Phi_c(m)$ or, in the limit of large $m$, for $\Phi>\Phi_c$. Furthermore, if $\Phi_c$ exists, we investigate whether $\Phi_c/2\pi$ is a rational number. We also look for a qualitative characterization of the transition occurring at $\Phi_c(m)$. To address these questions in a well-defined setting, we choose to consider for an isotropic flux with the magnetic field oriented along the diagonal of the cubic lattice, and with isotropic hopping coefficients.

%Our problem: define it and how we tackle it
Due to the involved analytical structure of the problem, an exact mathematical expression of $\Phi_c(m)$ as a function of $m$ is not present in the literature, to the best of our knowledge. Numerical estimates of $\Phi_c$ are present in few works \cite{Hasegawa1990,FontanaPRB2021}, and are based on the exact diagonalization (ED) of the 3D model. The aim of this work is to numerically investigate the existence of the critical flux $\Phi_c(m)$ for general co-prime pairs $(m,n)$. By means of momentum space ED, we analyze the structure of isolated band touching points for various magnetic fluxes, by increasing progressively the value of $m$. Based on the sequence of critical fluxes obtained for different values of $m$, $\Phi_c(m)$ for $m \leq 60$, we argue in favour of an asymptotic finite value of $\Phi_c$ for large $m$, by keeping the ratio $m/n$ finite. Additionally, based on the same sequence of values, we formulate a conjecture for the value of $\Phi_c(m)$ to be used in the large-$m$ limit. The resulting value of $\Phi_c$, referred to as $\Phi_c^{{\rm (conj)}}$, yields a rational number, in agreement with the numerical estimate of $\Phi_c$ obtained from a fit using the values of $\Phi_c(m)$. We also characterize other relevant physical quantities and the momentum space bands structure of the model for large $m$, such as the ground state energies and the isolated band touching points.

The paper is organized as follows. In Sec \ref{3d_hofstdadter} we introduce the 3D Hofstadter model and show its diagonalization in momentum space. In Sec. \ref{phic_definitions} we define the spectral measures of the reduced Hamiltonian and how they are related to the possible definitions of the critical flux. In Sec. \ref{numerical_results} we present our numerical results on the band structure of the model for $m\geq2$. We discuss the scaling of the ground state energy, the band touching points position in momentum space and the numerical determination of the critical flux. In Sec. \ref{math_conjecture} we present our mathematical conjecture for the critical flux, based on our numerical results for $m\leq 60$. In Sec. \ref{conclusions} we summarize and present our conclusions. The various Appendixes contain numerical details and computations, and the relevant tables with the pairs $(m,n)$ analysed in the manuscript.

\section{\label{3d_hofstdadter}The Hofstadter model}
The real-space Hamiltonian of the Hofstadter model is
\begin{equation}
    H = -  t \, 
    \sum_{{\bm r} \, , \, \hat{j}}   c^{\dagger}_{{\bm r} + \hat{j}} 
    e^{i \theta_{j}\left({\bm r}\right)} 
    c_{{\bm r} } + \ \mathrm{H.c.} \, ,
    \label{hofst}
\end{equation}
where $t$ is the hopping amplitude, assumed equal in the three directions $x,y,z$, $c^\dagger_{\bm{r}}$ and $c_{\bm{r}}$ are the creation and annihilation operators, and we used the Peierls substitution 
\begin{equation}
    \theta_j(\bm{r})\equiv\int_{\bm{r}}^{\bm{r}+\hat{j}}\;\bm{A}(\bm{x})\cdot d\bm{x}
    \label{peierls}
\end{equation}
to take into account the effect of the external magnetic fields, defining the fluxes across all plaquettes of the cube \cite{Peierls1933,Grosso_Parravicini_2000}. We consider the case of isotropic commensurate magnetic fluxes 
\begin{equation}
    \Phi=\frac{2\pi m}{n},\qquad m,\;n\in\mathbb{N}
    \label{rational_flux}
\end{equation}
with $(m,n)$ co-prime integer pairs corresponding to a magnetic field $${\bf B}= \Phi \left(1,1,1\right)$$ in units of the magnetic flux quantum $\Phi_0=h/(2e)$. In two dimensions, this is a paradigmatic model for the study of commensurability effects \cite{HofstadterPRB1976}. In the following we focus on the 3D case, considering a cubic lattice with $V=L^3$ sites, hopping amplitude $t=1$ and periodic boundary conditions (PBC).

The cubic lattice can be decomposed in independent sub-lattices whose size depends, in general, on the gauge choice. This freedom can be employed to identify the gauge producing the smallest number of sub-lattices related to the magnetic flux $\Phi$. As discussed and reviewed in \cite{BurrelloJPHYSMATH2017}, for a 3D cubic lattice the minimal set of sub-lattices has dimension $n$, and a practical gauge choice in this context is the Hasegawa gauge \cite{Hasegawa1990}, defined as
\begin{equation}
    \bm{A}(\mathbf{x})=\Phi \cdot (0,x-y,y-x) \, .
    \label{hasegawa_gauge}
\end{equation}
Within this gauge, the Peierls phases are 
\begin{equation}    
    \theta_x(\bm{r})=0,\qquad\theta_y(\bm{r})=\Phi\bigg(x-y-\frac{1}{2}\bigg),\qquad\theta_z(\bm{r})=\Phi(y-x) \,,
    \label{peierls_phases}
\end{equation}
immediately observing that the $z$-coordinate is absent. Another interesting feature of the Hasegawa gauge is the explicit dependence only on the relative difference $x-y$, which indicates that the problem is effectively 1D in momentum space \cite{KunsztZeePRB1991}.

\subsection{\label{kspace_diagonalization_subsec}Momentum space diagonalization}
The 3D Hofstadter model can be solved in momentum space by introducing the magnetic translation group, exploiting the interplay between gauge and translational invariance in the system when a commensurate background magnetic field is present \cite{Landau1981,ZakIPRA1964,ZakIIPRA1964}. For the gauge choice in Eq. \eqref{hasegawa_gauge}, we can define the magnetic Brillouin zone as
\begin{equation}
    \text{MBZ}:\quad k_x\in\bigg[-\frac{\pi}{n},\frac{\pi}{n}\bigg],\;k_y\in\bigg[-\frac{\pi}{n},\frac{\pi}{n}\bigg],\;k_z\in\bigg[-\pi,\pi\bigg]
    \label{MBZ_definition} \, .
\end{equation}
The Hamiltonian is then expressed in terms of independent blocks known as magnetic bands. Each sub-lattice corresponds to a specific band, and each of these bands exhibits an $n$-fold degeneracy. We notice that when ${\bf k}\in\text{MBZ}$, the allowed values of ${\bf k}$ are $L/n^2$, and for each of them the associated matrix is of size $n\times n$. Consequently, we obtain $L/n$ eigenvalues, each one being degenerate $n$ times, matching the real-space dimensionality of the problem, as described, for example, in \cite{BurrelloJPHYSMATH2017}.

The form of Eq. \ref{hofst} in momentum space is
\begin{equation}
    H=-t\sum_{\mathbf{k}\in\text{MBZ}}\sum_{\hat{j},s}c^\dag_{s',\mathbf{k}}(T_{\hat{j}})_{s',s}e^{-i\mathbf{k}\cdot\hat{j}}c_{s,\mathbf{k}}+\text{H.c.}\equiv-t\sum_{\mathbf{k}\in\text{MBZ}}C^\dagger_{\mathbf{k}}\mathcal{H}(\mathbf{k})C_{\mathbf{k}}\,,
    \label{reduced_H_momentumspace}
\end{equation}
where $s$ labels the magnetic bands and the $n \times n$ matrices $T_{\hat{j}}$ are for $\hat{j}=\hat{x},hat{y},\hat{z}$ given by 
\begin{equation}
    T_{\hat{x}}=
    \begin{pmatrix}
    0 & 1 & 0 & 0\\
    0 & 0 & \ddots & 0\\
    0 & \cdots & 0 & 1\\
    1 & 0 & \cdots & 0
    \end{pmatrix},
    \qquad
    T_{\hat{y}}=e^{-\frac{i\Phi}{2}}
    \begin{pmatrix}
    0 & \cdots & 0 & \varphi_0\\
    \varphi_1 & 0 & \cdots & 0\\
    0 & \ddots & 0 & 0 \\
    0 & 0 & \varphi_{n-1} & 0
    \end{pmatrix},
    \qquad
    T_{\hat{z}}=
    \begin{pmatrix}
    \varphi_0 & 0 & \cdots & 0\\
    0 & \varphi_{n-1} & 0 & 0\\
    0 & 0 & \ddots & 0\\
    0 & \cdots & 0 & \varphi_1
    \end{pmatrix}
    \label{Tmatrices_defs}
\end{equation}
in the sub-lattice basis. To lighten the notation, in Eq. \eqref{Tmatrices_defs} we defined $\varphi_l=e^{i\Phi l}=e^{\frac{2\pi i m l}{n}}$, with $l=0,\ldots,n-1$. It is worth to notice that for $n=2$, the well-known $\pi-$flux case, the model hosts a Weyl semimetallic phase at half-filling \cite{Affleck-MarstonPRB1989,LaughlinZouPRB1990,Lepori2010,KetterleePRL2015}, notably preserving physical time-reversal and space inversion symmetries \cite{Lepori_Fulga_Trombettoni_BurrelloPRB2016}. When $n\neq2$ we generally have the explicit breaking of time-reversal symmetry dictated by the presence of the external magnetic field, mathematically reflected into the structure of the matrices $T_{\hat{j}}$, which are not invariant under the conjugate operation. Additionally, the real-space Hamiltonian in Eq. \eqref{hofst} has a chiral sub-lattice symmetry $c_{\mathbf{r}}\rightarrow(-1)^{x+y+z}c_{\mathbf{r}}$, which maps $\mathcal{H}\rightarrow-\mathcal{H}$. As a consequence, the model shows a symmetric single-particle energy spectrum.

The energy dispersion relations of the $n$ bands of the model can be obtained by diagonalizing the matrix $\mathcal{H}(\mathbf{k})$, whose general structure is
\begin{equation}
    \mathcal{H}(\mathbf{k})=
     \begin{pmatrix}
        D_0 & U_1 & 0 & \ldots & 0 & U_0^*\\
        U_1^* & D_{n-1} & U_2 & 0 & \ldots & 0\\
        0 & U_2^* & D_{n-2} & U_3 & 0 & \ldots\\
        \vdots & \ddots & \ddots & \ddots & \ddots & \ddots \\
        \vdots & \ddots & \ddots & \ddots & \ddots & U_{n-1}\\
        U_0 & 0 & \ldots & \ldots & U_{n-1}^* & D_1\\
        \label{tridiagonal_matrix}
    \end{pmatrix}\,,
\end{equation}
where
\begin{equation}
    D_j(k_z)=e^{-ik_z}\varphi_j+\text{H.c.}=2\cos{(k_z-j\Phi)},\qquad U_j(k_x,k_y)=e^{-ik_x}+e^{i(\Phi/2+k_y)}\varphi_j^*\,.
\end{equation}
To summarize, the spectral problem of the 3D Hofstadter model with isotropic commensurate fluxes is equivalent to a family of $n\times n$ matrices parametrized by the three continuous parameters $\bf{k}\in\text{MBZ}$. The periodic tridiagonal structure obtained in Eq. \eqref{tridiagonal_matrix} at fixed values of ${\bf k}$ corresponds to a periodic Jacobi matrix \cite{Molinari1997,Molinari2003}. For this class of matrices, the Hamiltonian level dynamics can be related to a $2\times2$ matrix known as the transfer matrix, as already pointed out by Hofstadter in his celebrated work for the 2D model \cite{HofstadterPRB1976}. Theorems and results for energy bands and gaps measures in the case of real Jacobi matrices have been extensively discussed \cite{Last1992,Last1993,Korotyaev2003,Shamis2011}, and there exists a criterion to establish if the measure of union of the gaps in the spectrum has vanishing measure, depending on the elements on the first diagonal of the Jacobi matrix \cite{Korotyaev2003}. The extension of these results to certain classes of complex Jacobi matrices has also been recently studied recently in the mathematical literature \cite{Swiderski2020}.

\section{\label{phic_definitions}Spectral measures and critical flux definitions}
Before showing our numerical results and analytical conjecture for the critical flux, we define the fundamental quantities we look at in the analysis of the bands structure of the Hofstadter model. As previously stated, the reduced Hamiltonian $\mathcal{H}({\bf k})$ has $n$ degenerate bands, labelled as $B_j({\bf k})$, $j=0,\ldots,n-1$. For any of these bands we define
\begin{equation}
    \min{B_j({\bf k})}=\epsilon_j({\bf k}),\qquad
    \max{B_j({\bf k})}=E_j({\bf k}),\qquad j=1,\ldots,n\,,
    \label{min_max_bands_definition}
\end{equation}
and the energy range of the $j$-band is $\sigma_j=[\epsilon_j,E_j]$. The union $\cup_j\sigma_j\equiv\sigma(\mathcal{H})$ is defined as the spectrum of the 3D Hofstadter problem at the fixed flux pair $(m,n)$. For any pair of consecutive bands $B_j$, $B_{j+1}$ we can have
\begin{equation}
    B_j\cap B_{j+1}=\emptyset\qquad\text{or}\qquad B_j\cap B_{j+1}\equiv\mathcal{O}_{j,j+1}\neq\emptyset \,,
\end{equation}
and in the first case the bands do not overlap, while in the second case they do, and we denote with $\mathcal{O}_{j,j+1}$ their intersection. In the 2D case, we can also define the set of the gaps as
\begin{equation}
    G(\mathcal{H})\equiv[\min\sigma(\mathcal{H}),\max\sigma(\mathcal{H})]/\sigma\equiv\cup_j[E_j,e_{j+1}],\qquad j=1,\ldots,n-1\,,
\end{equation}
since it is well known that $G(\mathcal{H})\neq\emptyset$ \cite{HofstadterPRB1976,Harper&RoyPRB2014}. In this respect, the 3D is generically different, as the gap set's measure depends on the orientation of the magnetic fields $\bf{B}$. In the case of a cubic lattice, for directions of the field that are not the high-symmetry crystallographic ones the spectrum is gapful \cite{KoshinoPRB2003}. For ${\bf B}\propto(1,1,1)$, which is the case considered here, the spectrum is \textit{gapless} for any co-prime pair $(m,n)$ \cite{Hasegawa1990,Hasegawa1992}. Nonetheless, we can have $B_j\cap B_{j+1}=\emptyset$ for some values of $j$, corresponding to bands touching in isolated points: if a co-prime pair realizes this situation, the measure spectrum has a peculiar structure \cite{Hasegawa1990}, and can be expressed as the union of three non-overlapping sets
\begin{equation}
    \sigma(\mathcal{H})=\sigma_1\cup\sigma_2\cup\sigma_3,\qquad \sigma_1\equiv\bigcup_{j=1}^{m}B_j,\qquad\sigma_2\equiv\bigcup_{j=1}^{n-2m}B_{j+m},\qquad\sigma_3\equiv\bigcup_{j=1}^{m}B_{j+n-m} \,.
    \label{hasegawa_spectral_measure}
\end{equation}
With this spectral measure, and due to the symmetry properties of $\mathcal{H}({\bf k})$, to understand if there is overlap between all the degenerate groups of bands we can simply look at $\mathcal{O}_{m,m+1}$, i.e. the intersection between $\sigma_1$ and $\sigma_2$. If instead the co-prime pair does not realize this situation, all the degenerate blocks of bands overlap. 

Given these considerations, we can define the critical flux $\Phi_c (m)$ at fixed $m$ as the magnetic flux at which we observe the transition from disjoint blocks of degenerate bands to overlapping bands. In any case, it is important to notice that $\Phi/2\pi$ is not irrational, and parametrized by co-prime integer ratios $\Phi\propto m/n$. At fixed $m$, therefore, we have $\Phi_c(m)\propto n_c^{-1}$, allowing for the identification of the corresponding critical integer $n_c$. It is worth reminding that small fluctuations of $\Phi$ result in significant variations of the co-prime pairs $(m,n)$ \cite{HofstadterPRB1976,Hasegawa1990,KunsztZeePRB1991}, meaning that arbitrary close values of $\Phi$ could be represented by co-prime pairs that are apparently unrelated and ``far away" from each other. To generalize the definition of $\Phi_c$ towards the limit $n_c\rightarrow\infty$, and to identify a regular and well-behaved succession of co-prime pairs such that, in units of $2\pi$,
\begin{equation}
    \frac{m_1}{n_{c,1}},\;\frac{m_2}{n_{c,2}},\;\cdots,\;\frac{m_\ell}{n_{c,\ell}}\;\xrightarrow{\ell\rightarrow\infty}\;\frac{\Phi_c}{2\pi}=\frac{m_\infty}{n_{c,\infty}}
    \label{succession_ratios}\,,
\end{equation}
we need to give a more precise meaning to the critical integer $n_{c,\ell}$. 

We propose to look to four different definitions, for any integer $m$:
\begin{enumerate}
    \item the element $n_{c,m}$ is identified as the first co-prime integer such that the bands $B_m$, $B_{m+1}$ of the model \textit{do overlap}:
    \begin{equation}
        \{n_{c,m}\in\mathbb{N}|\;\text{gcd}(m,n)=1,\;\mathcal{O}_{m-1,m}=\emptyset\;\text{and}\;\mathcal{O}_{m,m+1}\neq\emptyset\}
        \label{nc_succession_def}\,;
    \end{equation}
    
    \item the element $n'_{c,m}$ is identified as the last co-prime integer immediately before the bands overlap. In this case, the bands $B_m$, $B_{m+1}$ \textit{do not overlap}:
    \begin{equation}
        \{n'_{c,m}\in\mathbb{N}|\;\text{gcd}(m,n)=1,\;\mathcal{O}_{m,m+1}=\emptyset\;\text{and}\;\mathcal{O}_{m+1,m+2}\neq\emptyset\}
        \label{n'c_succession_def} \,;
    \end{equation}
    
    \item the element $\tilde{n}_{c,m}$ defined as the first \textit{non-co-prime} integer value for which the bands \textit{do overlap}:
    \begin{equation}
        \{\tilde{n}_{c,m}\in\mathbb{N}|\;\mathcal{O}_{m-1,m}=\emptyset\;\text{and}\;\mathcal{O}_{m,m+1}\neq\emptyset\}
        \label{ntilde_succession_def}\,;
    \end{equation}
    
    \item the element $\tilde{n}'_{c,m}$ defined as the last \textit{non-co-prime} integer value for which the bands \textit{do not overlap}:
    \begin{equation}
        \{\tilde{n}'_{c,m}\in\mathbb{N}|\;\mathcal{O}_{m,m+1}=\emptyset\;\text{and}\;\mathcal{O}_{m+1,m+2}\neq\emptyset\}
        \label{ntilde_prime_succession_def}\,.
    \end{equation}
    
\end{enumerate}

With the last two definitions we simply want to include critical pairs that would be neglected by the co-prime condition, and whose exclusion may give rise to a wrong estimation of the real values of the critical flux $\Phi_c$. Due to these definitions, it is immediate to verify that $\tilde{n}_{c}\leq n_c$ and $\tilde{n}'_c \geq n'_c$, at fixed $m$. As will be shown in the next Section, these different definitions produce compatible results and agree in the large-$m$ limit. Among the aforementioned four definitions, somehow the most practical and intuitive is the last one, that we use to define the critical value at fixed $m$:
\begin{equation}
    \Phi_c(m) = 2 \pi \cdot \frac{n}{\tilde{n}'_{c,m}}\,.
\end{equation}
When there is no risk of ambiguity, we will use either $n_c(m)$ or simply $n(m)$ instead of $\tilde{n}'_{c,m}$.

In the next Section, we present numerical results for the critical flux $\Phi_c(m)$ obtained from the diagonalization of the matrix $\mathcal{H}({\bf k})$. Based on the co-prime pairs extracted with this procedure, we provide our estimate for $$\Phi_c = \lim_{m\to \infty} \Phi_c(m) = \lim_{m \to \infty} 2 \pi \cdot \frac{m}{n(m)}$$ in Section \ref{math_conjecture}.

\section{\label{numerical_results}Numerical results}
We present the results obtained with large ED of the Hamiltonian in Eq. \eqref{reduced_H_momentumspace} for different co-prime pairs $(m,n)$ parametrizing the magnetic flux $\Phi$. Apart from Subsection \ref{m=1_results}, where we briefly remind the known results for $m=1$, we will consider integers $m\geq2$. Besides the critical flux identification, we aim to characterize, at fixed $m$, the energy features of the model as a function of $n$.

\subsection{\label{m=1_results}Summary of results for $m=1$}
The properties of the 3D Hofstadter model with flux $\Phi=2\pi/n$ have recently been studied as a function of $n\in\mathbb{N}$ \cite{FontanaPRB2021}. By looking at the two definitions given by Eqs. \eqref{nc_succession_def} and \ref{n'c_succession_def}, for this particular case we have
\begin{equation}
    n_c=\tilde{n}_c=8,\qquad n'_c=\tilde{n}'_c=7\,.
    \label{m=1_n_critical}
\end{equation}
As a general feature, for $n<n_c$ the density of states (DOS) of the system shows isolated zeros, associated to Weyl points separating the lowest band from the others. The corresponding filling at the Weyl points is $\nu=n^{-1}$, generalizing the result for $n=2$ to higher integers \cite{Affleck-MarstonPRB1989,Hasegawa1990,Lepori2010}. In the opposite regime $n>n_c$, the DOS does not show any zero. As a consequence, there are values of the chemical potential $\mu$ for which the system is in a multi-band metallic state, and the corresponding Lifshitz transitions between these states and single band metallic phases can be identified.

The lowest energy bands extrema, denoted by $\epsilon_0$ and $E_0$, are respectively the ground state energy and the Weyl energy, and scales with $n$ with the power laws $\epsilon_0\sim n^{-0.8}$, $E_0\sim n^{-2.8}$ \cite{FontanaPRB2021}.

\subsection{\label{energy_bands}Energy bands structure and inversion points for $m\geq 2$}
\begin{figure}[h!]
    \centering
    \includegraphics[width=0.2\linewidth]{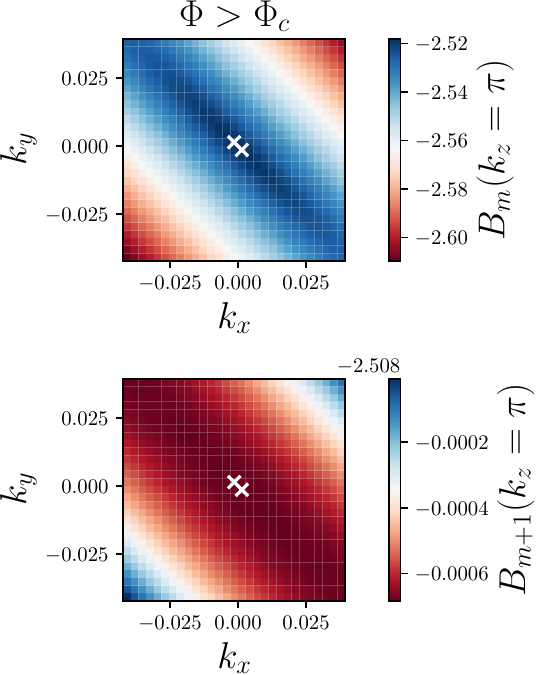}
    \qquad
    \includegraphics[width=0.2\linewidth]{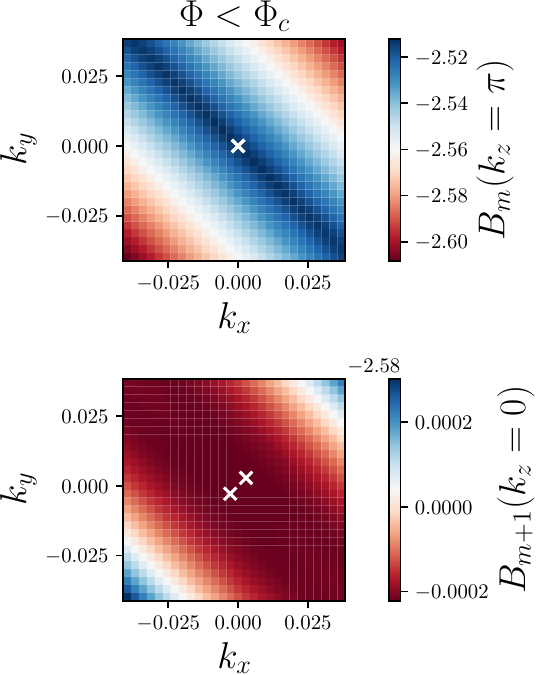}
    \qquad
    \includegraphics[width=0.2\linewidth]{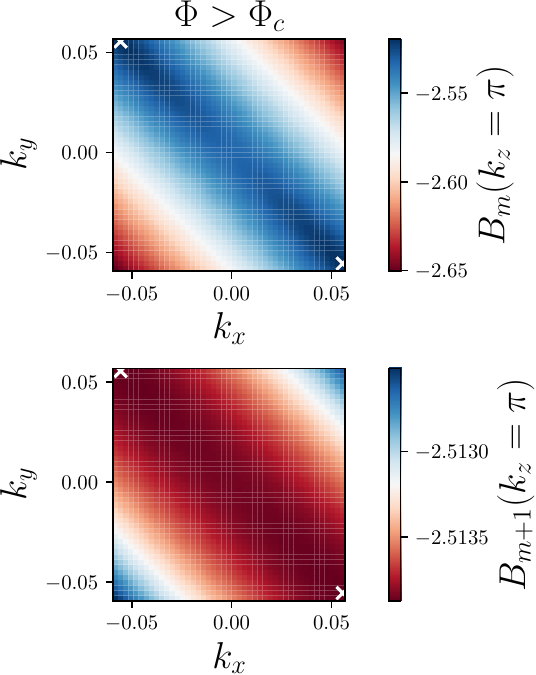}
    \qquad
    \includegraphics[width=0.2\linewidth]{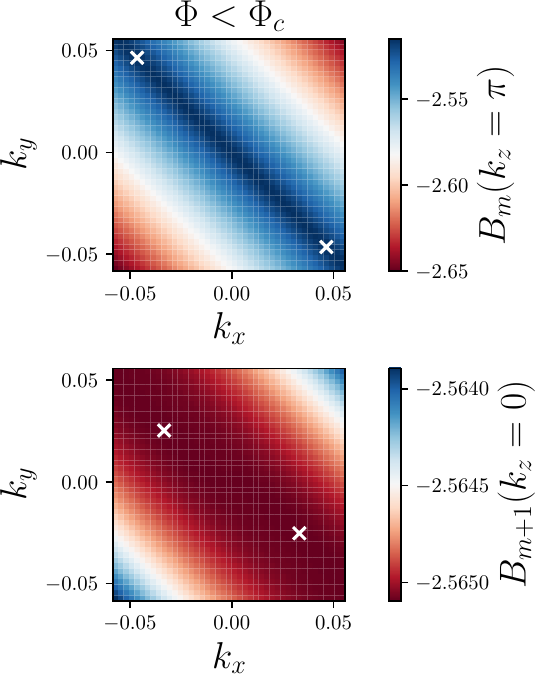}
    \caption{Color plots of $B_m({\bf k}),\;B_{m+1}({\bf k})$ in the $(k_x,k_y)$ plane for $m=10$, $n=77,\;79$ (two leftmost columns) and $m=7$, $n=54,\;55$ (two rightmost columns) with $k_z=0,\pi$ as written in the color bar labels. White cross markers represent ${\bf k}_{\text{min}}$ in the $(k_x,k_y)$ plane, whose explicit coordinates are reported in Table \ref{table1:example_cases_aroundphi_C}.}      
    \label{4m_bands_momentumspace_plots}
\end{figure}
For values of up to $m=10$, we numerically diagonalize the Hamiltonian by spanning the full MBZ with discretization of size $\delta {\bf k}=O(10^{-3})$ in all directions, and investigate how the bands $B_m({\bf k})$, $B_{m+1}({\bf k})$ behave. We find that for values of $\bf k$ in the planes 
\begin{equation}
    \Pi_{\bar{k}_z}=\{{\bf k}\in\text{MBZ}|k_z=\bar{k}_z\},\qquad \bar{k}_z=0,\pi\,,
    \label{planes_reducedMBZ}
\end{equation}
we are able to extract the minimum of the lower band $\epsilon_m$ and explore the energy interval $[E_m,\epsilon_{m+1}]$ (see details in Appendix \ref{numerical_details_bands}). If the co-prime pair is such that $\Phi>\Phi_c$, the overlap $\mathcal{O}_{m,m+1}=\emptyset$ and $\epsilon_m=\epsilon_0$ \cite{Hasegawa1990}, allowing us to extract the ground state energy $\epsilon_0$.

We characterize the band structure around the band touching points ${\bf k}\approx{\bf k}_W$ by varying the magnetic flux $\Phi$. As a general pattern, almost independently of the co-prime pairs $(m,n)$, we observe that when $\Phi$ crosses the critical value from above, i.e. $\Phi\searrow\Phi_c$, the value $E_m$ remains the same, while the energy $\epsilon_{m+1}$ of degenerate central block of bands is lowered, giving rise to the non-trivial overlap when $\Phi<\Phi_c$. At the same time, we observe an abrupt change in the $z$-component of the momentum ${\bf k}_{\text{min}}$ associated to $\epsilon_{m+1}$, with associated values of $(k_x,k_y)$ that depends on the parity of $m$. We have
\begin{equation}   
    \Phi>\Phi_c\rightarrow\Phi<\Phi_c\qquad\Rightarrow\qquad {\bf k}_{\text{min}}=(0,0,\pm\pi)\rightarrow{\bf k}_{\text{min}}=(0,0,0),\qquad m\;\text{even}\,,
    \label{jump_momentum_kmin_even_m}
\end{equation}
\begin{equation}    
    \Phi>\Phi_c\rightarrow\Phi<\Phi_c\qquad\Rightarrow\qquad {\bf k}_{\text{min}}=\bigg(\mp\frac{\pi}{n},\pm\frac{\pi}{n},\pm\pi\bigg)\rightarrow{\bf k}_{\text{min}}=(\pm\bar{k}_x,\mp\bar{k}_x,0),\qquad m\;\text{odd}\,,
    \label{jump_momentum_kmin_odd_m}
\end{equation}
where $\bar{k}_{x}\in\text{MBZ}$ is determined numerically, up to the precision with which we span the MBZ. For even $m$, the minimum is always in the origin of the $x-y$ plane in momentum space, in both the regimes of $\Phi$. On the other hand, for odd $m$ the bands extrema are located at opposite corners of the MBZ for $\Phi>\Phi_c$, and they are shifted towards the origin along the line $k_x=-k_y$ when we cross the critical value, as reported in Eq. \eqref{jump_momentum_kmin_odd_m}. The jump in $k_z$ is a common feature of all $m$, suggesting the presence of a band inversion point. We report in Appendix \ref{jacobi_inversion_points_App} the form of the Jacobi matrix $\mathcal{H}({\bf k})$ in correspondence of the momenta of Eq. \eqref{jump_momentum_kmin_even_m}, \eqref{jump_momentum_kmin_odd_m}. 

We refer to Table \ref{table1:example_cases_aroundphi_C} for examples with specific values of $(m,n)$, and to Fig. \ref{4m_bands_momentumspace_plots} for two examples with even and odd $m$. This provides another way of extracting the critical flux, based on physical arguments: by looking at the bands structures of $B_m({\bf k}),\; B_{m+1}({\bf k})$ in the reduced MBZ for various $\Phi$, the critical flux $\Phi_c$ is the one for which there is a band inversion point in the $(m+1)-$band along the $z$-direction.
\begin{table}[h]
    \setlength{\tabcolsep}{10pt}
    \begin{center}
        \begin{tabular}{cc|cccc}
            \hline 
            \hline
            $m$ & $n$ & $E_m$ & ${\bf k}_{\text{max}}$ & $\epsilon_{m+1}$ & ${\bf k}_{\text{min}}$ \\  
            \hline
            4 & 29 & $-2.52$ & $(0,0,\pi)$ & $-2.50$ & $(0,0,\pi)$\\
              & 31 & $-2.52$ & $(0,0,\pi)$ & $-2.53$ & $(0,0,0)$\\
            6 & 43 & $-2.52$ & $(10^{-3},-10^{-3},\pi)$ & $-2.51$ & $(10^{-3},-10^{-3},\pi)$\\
              & 47 & $-2.52$ & $(0,0,\pi)$ & $-2.56$ & $(-3\cdot10^{-3},3\cdot10^{-3},0)$\\
            10 & 77 & $-2.52$ & $(-10^{-3},10^{-3},\pi)$ & $-2.51$ & $(-10^{-3},10^{-3},\pi)$\\
              & 79 & $-2.51$ & $(0,0,\pi)$ & $-2.58$ & $(2\cdot10^{-3},2\cdot10^{-3},0)$\\ 
            \hline
            5 & 38 & $-2.52$ & $(\mp0.077,\pm0.077,\pi)$ & $-2.50$ & $(\mp0.077,\pm0.077,\pi)$\\
              & 39 & $-2.52$ & $(\mp0.011,\pm0.011,\pi)$ & $-2.54$ & $(\mp0.040,\pm0.040,0)$\\
            7 & 54 & $-2.52$ & $(\mp0.055,\pm0.055,\pi)$ & $-2.51$ & $(\mp0.055,\pm0.055,\pi)$\\
              & 55 & $-2.51$ & $(\mp0.033,\pm0.033,\pi)$ & $-2.56$ & $(\mp0.046,\pm0.046,,0)$\\  
             
            \hline
            \hline
        \end{tabular}
    \end{center}
    \caption{Examples of values of energies $E_m$, $\epsilon_{m+1}$ for various co-prime pairs crossing the critical value $\Phi_c(m)$, alongside their position in momentum space. Pairs above the horizontal line within the table have even $m$, while the remaining ones have odd $m$. The MBZ is spanned in steps of size $\delta {\bf k}=O(10^{-3})(1,1,1)$ for all the considered pairs $(m,n)$.}
    \label{table1:example_cases_aroundphi_C}
\end{table}

According to these numerical observations, we are able restrict further the exploration of the MBZ in the ED algorithm, and speed up the numerics for large values of $n$. To extract the ground state energy $\epsilon_0$ and the bands touching points we explore the planes $\Pi_{\bar{k}_z}\subset\text{MBZ}$, while for the determination of $\Phi_c$ solely we can limit to the bisector $k_x=-k_y$ of $\Pi_{\bar{k}_z}$. In both cases, we have to take into account the corresponding uncertainty on the energies, which is higher if compared with the ED obtained with the full MBZ. For completeness, we report in Appendix \ref{table_energies_234digits} the energies $E_m$ and $\epsilon_{m+1}$ rounded to the second, third and fourth significant digits, showing how different levels of precision in the energies may lead to different estimates of the critical integers. Notably, when the entire MBZ is explored, we typically reach a maximum size $L \sim 500$, which in turn fixes the grid of the MBZ. The second column of the second and third tables in Appendix \ref{table_energies_234digits} are identical, while the second column of the first table (pertaining to the rounding of energies to the second digit) differs from the corresponding ones in the other tables only for $m=25$. This demonstrates the robustness of our results across the errors committed in the determination of the energy eigenvalues. 

From now on, unless differently specified in the text, we round the relevant energies to four decimal digits and use the corresponding integers as critical $n_c$ values.

\subsection{Ground state energy scaling and Weyl energies}
In addition to the momentum space characterization of the central bands $B_m({\bf k})$ and $B_{m+1}({\bf k})$, we characterize the scaling of the ground state energy $\epsilon_0=\min{\sigma(\mathcal{H})}$ and the band touching energies $\epsilon_{m+1}$ as a function of the magnetic flux $\Phi$. 
\begin{figure}
    \centering
    \includegraphics[width=0.45\linewidth]{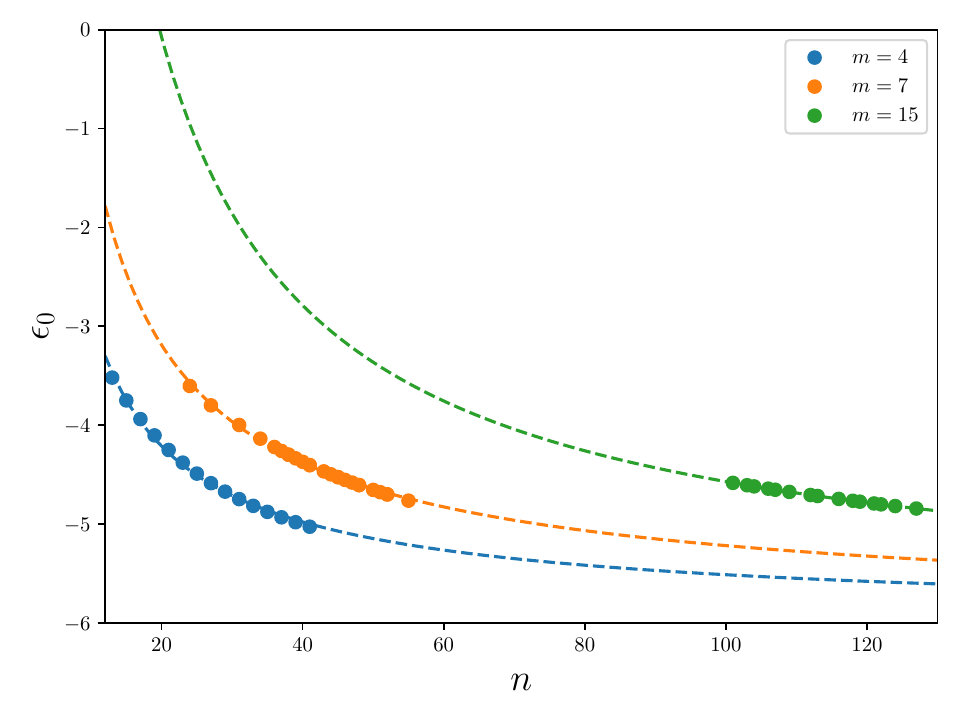}
    \qquad
    \includegraphics[width=0.45\linewidth]{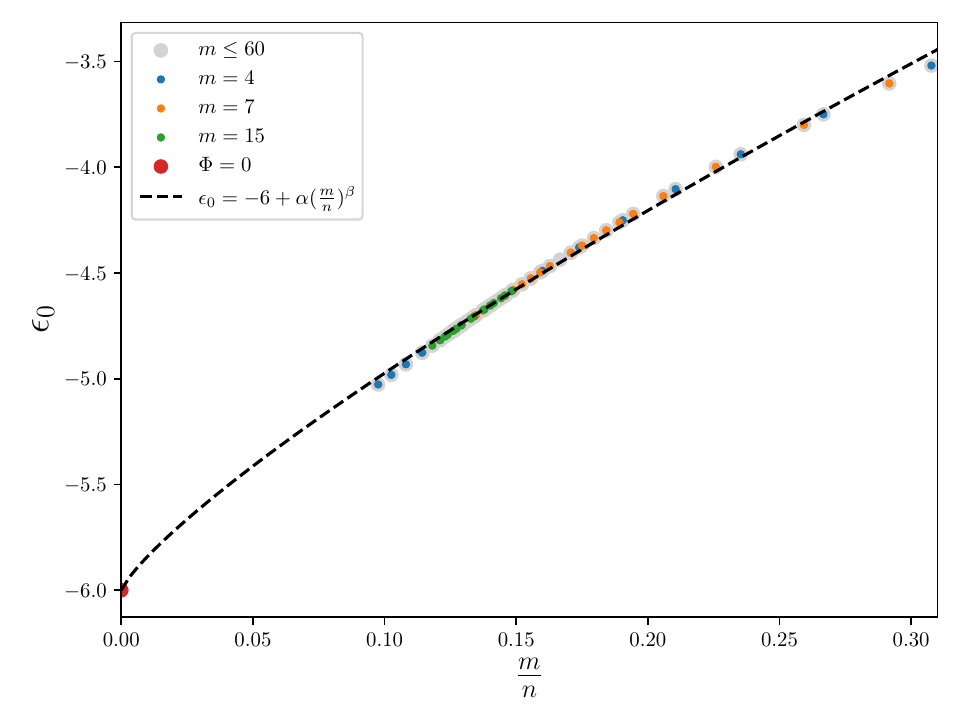}
    \caption{Left plot: $\epsilon_0$ vs $n$ at fixed values $m=4,7,15$. The dashed lines correspond to the fit function in Eq. \eqref{fit_e0_n_fixedm}. Right plot: Ground state energies $\epsilon_0$ for various fluxes $m/n=\Phi/2\pi$ up to $m=60$ (light grey circles). Data for $m=4,7,15$ are highlighted with different colors, and superimposed with the scaling fit function in Eq. \eqref{fit_e0_n_fixedm}. We also report the tight-binding point $\epsilon_0=-6$ ($t=1$).}
    \label{E0_plots}
\end{figure}

For what concerns the ground state energy, it is lowered as long as $\Phi$ is decreased, without any discontinuity when we cross the critical value $\Phi_c$. This is expected, since we progressively go towards the $\Phi\rightarrow0$ limit, recovering the standard tight-binding model in the absence of background magnetic fields, for which $\epsilon_0(\Phi=0)=-6$ when $t=1$. Our results indicate that $\epsilon_0(\Phi=0)=\min_\Phi{\epsilon_0}(\Phi)$.

The presence of $\Phi\neq0$ raises the ground state energy, in agreement with the $m=1$ case \cite{FontanaPRB2021}. In the left plot of Fig. \ref{E0_plots} we show $\epsilon_0$ as a function of $n$ at fixed values of $m$: ground state energies are on different curves, that we fit with the form
\begin{equation}
    \epsilon_0(n)=-6+\frac{a}{n^b},
    \label{fit_e0_n_fixedm}
\end{equation}
by taking into account the tight-binding limit $\epsilon_0(n\rightarrow\infty)=-6t$. We report the estimates of the parameters in Table \ref{table3:fit_params_e0_m}. The dependence on $m$ enters only in the multiplicative coefficient $a$, while, independently of $m$, the scaling relation is $\epsilon\sim n^{-0.8}$, consistent with the result obtained for $m=1$. 
\begin{table}[h]
    \setlength{\tabcolsep}{10pt}
    \begin{center}
        \begin{tabular}{c|cc}
            \hline 
            \hline
            $m$ & $a$ & $b$\\  
            \hline
            4 & 20.1(7) & 0.80(1)\\
            7 & 30.4(8) & 0.794(8)\\
            15 & 83.4(8) & 0.883(8)\\
            \hline
            \hline
        \end{tabular}
    \end{center}
    \caption{Fit parameters of Eq. \eqref{fit_e0_n_fixedm} obtained from the ED data in Fig. \ref{E0_plots} (right plot).}
    \label{table3:fit_params_e0_m}
\end{table}
We plot in Fig. \ref{E0_plots} (right plot) the various ground state energies for some of the flux ratios $m/n$ considered in our analysis, verifying the perfect collapse of the ED data as a function of the magnetic flux. We extrapolate the scaling function as
\begin{equation}
    \epsilon_0(\Phi)=-6t+\alpha\bigg(\frac{\Phi}{2\pi}\bigg)^\beta\,,
    \label{scaling_fit_functions}
\end{equation}
obtaining the general, i.e. independent of the co-prime pair $(m,n)$, estimates $\alpha=6.57(6)$, $\beta=0.806(5)$. The scaling function superimposed to the ED data is showed again in Fig. \ref{E0_plots} (right plot).

Regarding the band touching energy, due to the properties of the Hamiltonian for ${\bf k}\approx{\bf k}_W$ \cite{FontanaPRB2021} we do not observe a significant scaling of its location, as showed in Appendix \ref{EWeyl_determination}. From some of the entries in Table \ref{table1:example_cases_aroundphi_C},  it can be inferred that the location of the band touching point can be estimated or through $E_m$ or $\epsilon_{m+1}$, since these bands touch at isolated points. For progressively increasing values of $m$, and within the precision with which we sample the MBZ with the numerical ED, the location is almost constant and around the value $E_w=-2.5150(9)$ (see Appendix \ref{EWeyl_determination} for details) independently of the considered magnetic flux $\Phi>\Phi_c$.

\subsection{\label{critical_flux_estimation}Determination of the critical flux}
\begin{figure}
    \centering
    \includegraphics[width=0.45\linewidth]{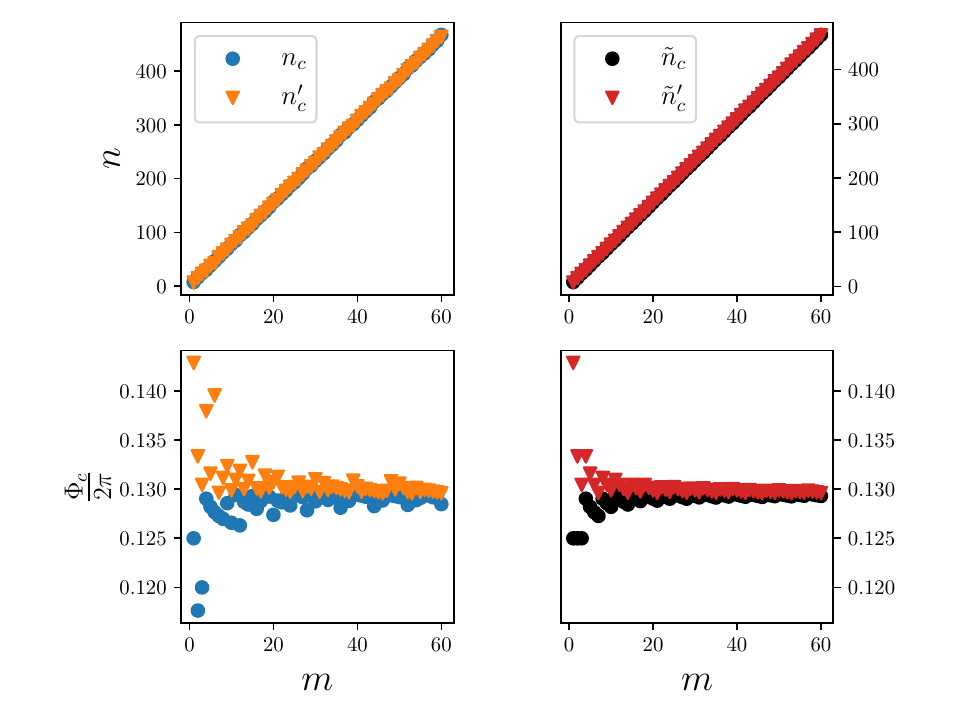}
    \qquad
    \includegraphics[width=0.45\linewidth]{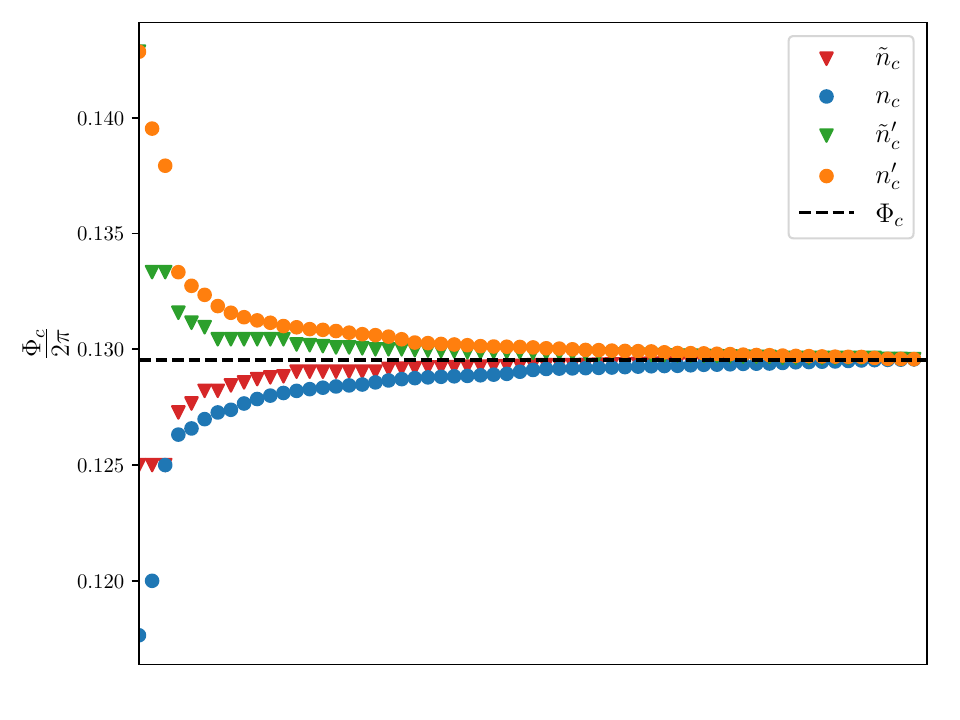}
    \caption{{\em Right plots:} Top subplots: all the critical pairs of Table \ref{tab:critical_couples} plotted as a function of $m$. Bottom subplots: $\Phi_c(m)/2\pi$ for the various $m\leq60$ for the four critical successions $n_c, n_c'$ (left subplot) and $\tilde{n}_c,\tilde{n}'_c$ (right subplot). The color code and markers are the same of the top subplots. {\em Left plot:} Critical fluxes for all the considered sequences, in ascending (for the not-primed sequences) or descending (for the primed sequences) order. This plot is not an indicator of how fast we reach the convergence to the $m\rightarrow\infty$ case.}
    \label{critical_sequences_fluxes_4subplots}
\end{figure}
We present the numerical estimation of the critical flux $\Phi_c(m)$ for various co-prime pairs $(m,n)$, with $m\geq 2$ up to $m=60$. By means of momentum space reduced ED in the planes $\Pi_{\bar{k}_z}$ introduced in Eq. \eqref{planes_reducedMBZ}, we extract the four sequences of critical integers $n$, explicitly reported in Table \ref{tab:critical_couples} (see Appendix \ref{table_critical_sequences}) and plotted in Fig. \ref{critical_sequences_fluxes_4subplots}. From the top subplots of Fig. \ref{critical_sequences_fluxes_4subplots}, it is evident the linear trend for all the four critical sequences identified in Section \ref{phic_definitions}. 

We plot as well all the critical fluxes $\Phi_c(m)/2\pi$, obtained with the same critical pairs, in the bottom subplots of Fig. \ref{critical_sequences_fluxes_4subplots}: as expected from their definitions, both $n_c',\;\tilde{n}_c'$ approach the critical flux from below, while the other two definitions $n_c,\;\tilde{n}_c$ from above. We observe a more regular trend by considering the two sequences $\tilde{n}_c,\;\tilde{n}'_c$, as they satisfy $|\tilde{n}_c-\tilde{n}'_c|=1$ at fixed $m$. This is particularly clear by looking at the bottom right subplot of Fig. \ref{critical_sequences_fluxes_4subplots}. For this reason, from now on we focus on the determination of $\Phi_c(m)$ using these two sequences, and comment on the estimate of $\Phi_c(m)$ estimated through $n_c'$ and $n_c$ in Appendix \ref{overlaps_analysis}.

\begin{figure}
    \centering
    \includegraphics[width=0.4\linewidth]{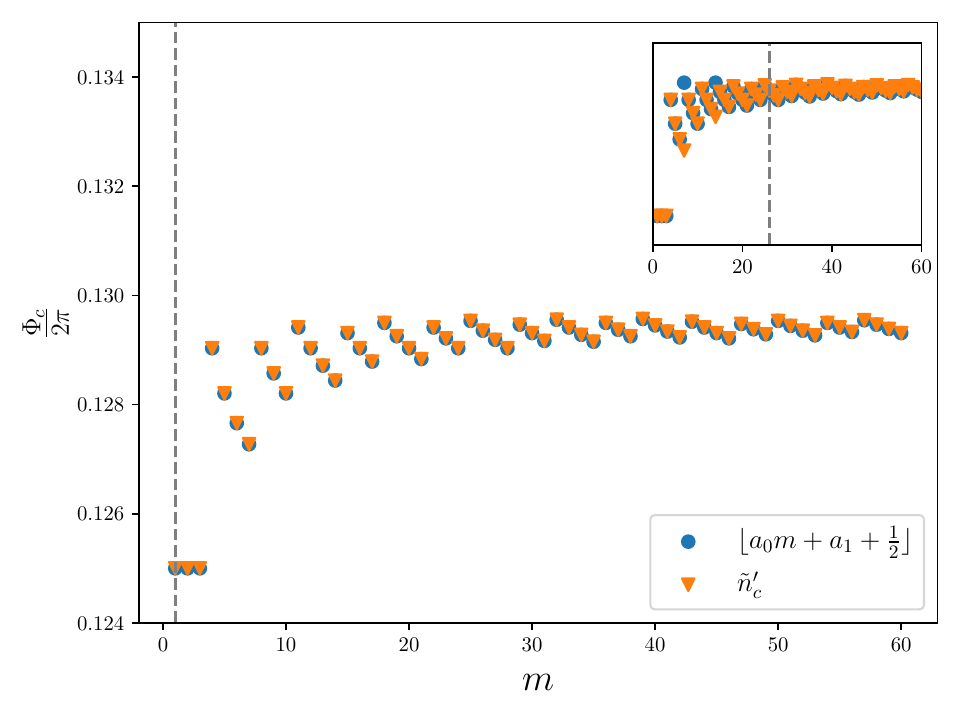}
    \qquad
    \includegraphics[width=0.4\linewidth]{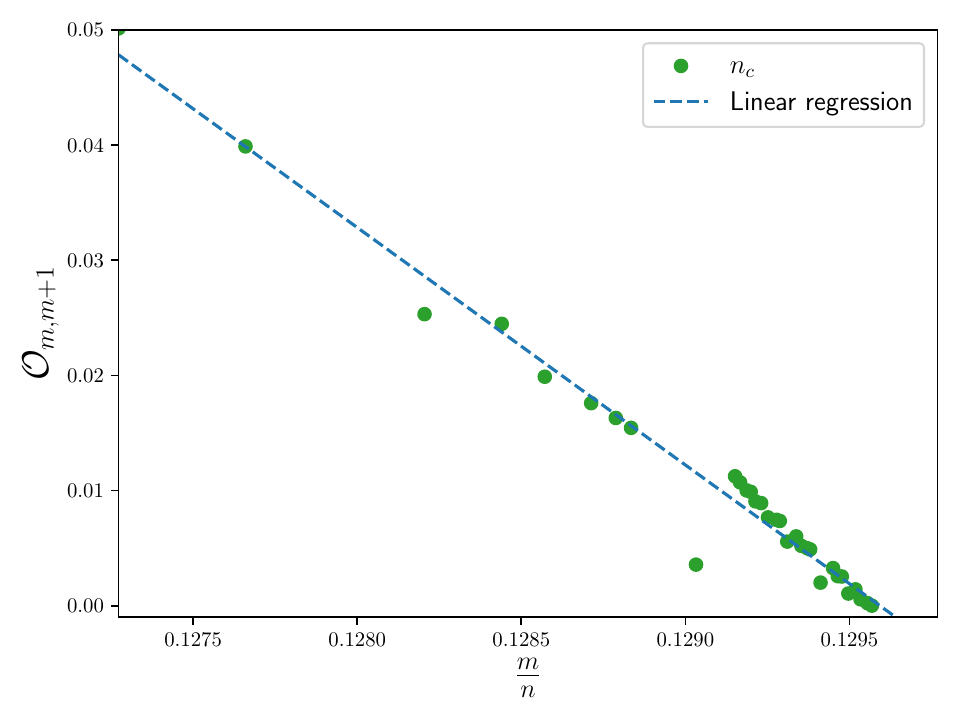}
    \caption{Left plot: Critical fluxes $\Phi_c(m)/2\pi$ for $\tilde{n}'_c$ together with the estimates of Eq. \eqref{phi_c_RHU definition}. The parameters $a_0$, $a_1$ of Eq. \eqref{phi_c_RHU definition} are obtained by discarding the values of $m$ below the threshold represented by the grey dashed vertical line: in the main plot and the inset, respectively, $m\geq 1$ and $m\geq 25$. Right plot: Bands overlaps as a function of $m/n$ for the co-prime pairs $(m,\tilde{n}_c)$ listed in Table \ref{tab:critical_couples}, superimposed with the straight line $\mathcal{O}_{m,m+1}=m(\Phi/2\pi)+q$.}
    \label{phi_c_fit_tilde data_plots}
\end{figure}

Based on these considerations, we perform a linear regression on the $\tilde{n}_c$, $\tilde{n}'_c$ data with straight lines $h(m) \equiv a_0m+a_1$, obtaining the results
\begin{equation}
    \tilde{n}_c:\quad a_0=7.718(2),\;a_1=0.49(8),\qquad\qquad \tilde{n}'_c:\quad a_0=7.718(2),\;a_1=-0.50(8)\,,
    \label{linear_fit_parameters_tildes}
\end{equation}
consistent with the properties of the critical sequences. However, since $m,n\in\mathbb{N}$, when we compare this estimate with the data of the critical flux $\Phi_c(m)$ we have to round the values obtained with $h(m)$. We choose to round using the round-half up function, i.e. half-way values of $h(m)$ are always rounded up to the corresponding integer value:
\begin{equation}
    h(m)\in\mathbb{R}\quad\rightarrow\quad \bigg\lfloor h(m)+\frac{1}{2}\bigg\rfloor\in\mathbb{N}\,.
\end{equation}
With this definition, the estimated critical flux as a function of $m$ is
\begin{equation}
    \frac{\Phi_c(m)}{2\pi}=\frac{m}{\lfloor a_0 m+a_1+\frac{1}{2}\rfloor}\,.    
    \label{phi_c_RHU definition}
\end{equation}
We plot this function with momentum ED data in Fig. \ref{phi_c_fit_tilde data_plots}.
Regarding the estimation of the parameters $a_0,\;a_1$, we extrapolate them for all considered the values of $m$, as we observe an excellent agreement with the numerical data. However, we observe that if we progressively discard data at low $m$ and fit only for values of $m$ above a given threshold, few isolated points are not consistent with our estimate (inset in left plot of Fig. \ref{phi_c_fit_tilde data_plots}).

By considering Eq. \eqref{phi_c_RHU definition} as the form of the critical flux $\Phi_c(m)$ as a function of $m$, we simply observe that
\begin{equation}
    \lim_{m\rightarrow\infty}\frac{\Phi_c(m)}{2\pi}=\frac{1}{a_0} = \Phi_c
    \label{phi_c_a0_relation_asymptotic}
\end{equation}
is the asymptotic estimate of the critical flux. For the two sequences $\tilde{n}_c$, $\tilde{n}'_c$ we estimate the same $a_0$, therefore we get
\begin{equation}
    \tilde{n}_c,\;\tilde{n}'_c:\quad \Phi_c=0.1296(1).
    \label{phi_c_estimates_twotilde_sequences}
\end{equation}
This is a unique and well defined quantity, as it coincides for both the sequences, consistent with the chosen energies uncertainties and within the precision of the parameters $a_0,\;a_1$. The critical co-prime pairs in Table \ref{tab:critical_couples} closest to this value are
\begin{equation}
    \frac{m}{\tilde{n}_c}=\frac{39}{301},\qquad\qquad\bigg|\frac{\Phi_c}{2\pi}-\frac{m}{\tilde{n}_c}\bigg|=2\cdot10^{-6},
    \label{tilde_nc_phi_c_estimate}
\end{equation}
\begin{equation}
    \frac{m}{\tilde{n}'_c}=\frac{46}{355},\qquad\qquad\bigg|\frac{\Phi_c}{2\pi}-\frac{m}{\tilde{n}'_c}\bigg|=7\cdot10^{-6}.
    \label{tilde_nc_prime_phi_c_estimate}
\end{equation}

On the same line, Eq. \eqref{phi_c_estimates_twotilde_sequences} is also consistent with the estimated value that can be obtained from the sequence $\tilde{n}_c$ through the analysis of the overlaps $\mathcal{O}_{m,m+1}$, introduced in Section \ref{phic_definitions}. In this case $\Phi\nearrow\Phi_c$, and we look at the functional dependence on $m$ of the quantity $\mathcal{O}_{m,m+1}\equiv|\epsilon_{m+1}({\bf k})-E_m({\bf k})|$ as long as we go towards $\Phi_c$. The results obtained for several co-prime pairs $(m,\tilde{n}_c)$ are plotted in Fig. \ref{phi_c_fit_tilde data_plots} (right plot). In the explored region, the overlaps close almost linearly as a function of the flux ratio $m/n$. We therefore perform a linear fit of the form  
\begin{equation}
    \mathcal{O}_{m,m+1}=m\frac{\Phi}{2\pi}+q,
    \label{linear_regression_overlaps}
\end{equation}
obtaining $m=-20.6(6)$ and $q=2.66(7)$. The critical flux $\Phi_c$ is extrapolated as
\begin{equation}
    \mathcal{O}_{m,m+1}=0\qquad\Rightarrow\qquad \Phi_c=-\frac{2\pi q}{m}
\end{equation}
which corresponds to $m/n_c=0.129(1)$, in agreement with Eq. \eqref{phi_c_estimates_twotilde_sequences}. We finally point out that the other sequences of integers, i.e. $n_c$ and $n_c'$, are consistent with the estimated $\Phi_c$, as can be seen from Fig. \ref{critical_sequences_fluxes_4subplots} (right plot), with all the numerical details left in Appendix \ref{overlaps_analysis}.

\section{\label{math_conjecture}Two conjectures and the asymptotic critical flux}
Based on the extrapolated form of the critical sequences, we propose two mathematical conjectures to establish the critical integer $n(m)$ as a function of $m$. Specifically, we focus on the sequence $\tilde{n}'_c$, but our approach can be translated to the other sequences as well.

\begin{figure}
    \centering
    \includegraphics[width=0.4\linewidth]{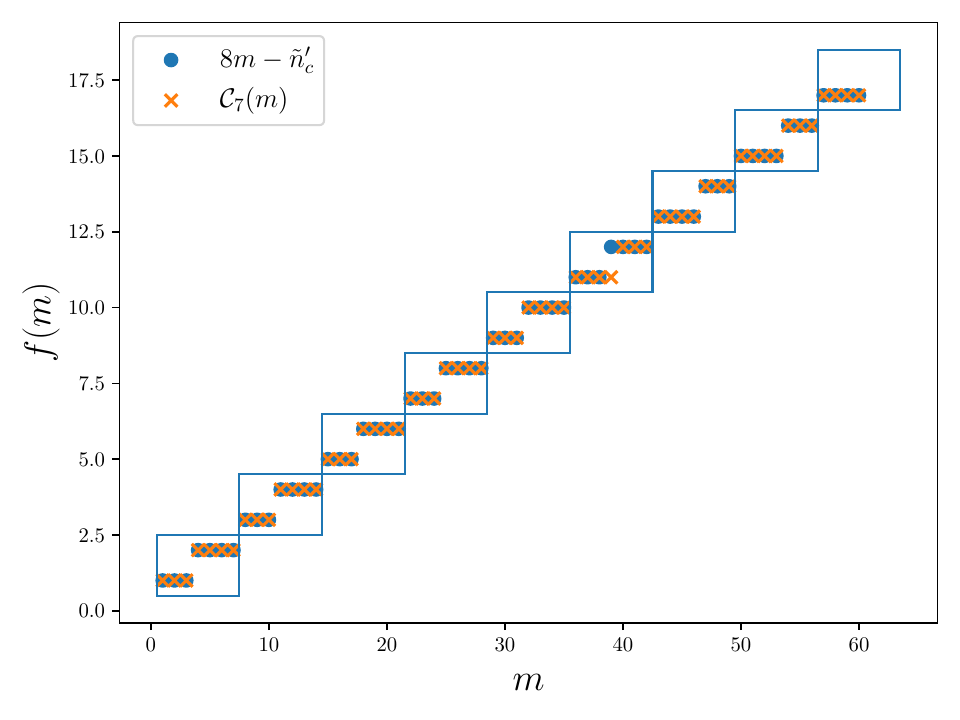}
    \qquad
    \includegraphics[width=0.4\linewidth]{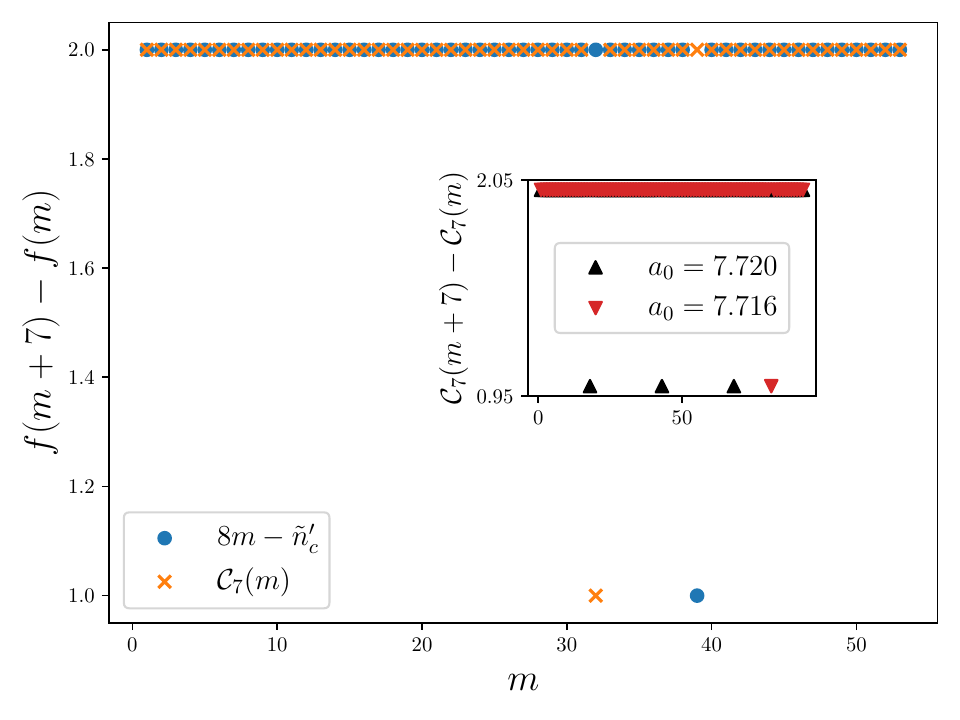}
    \caption{Left plot: Values of $\mathcal{C}_7(m)$ obtained with the mathematical conjecture \eqref{first_conjecture} superimposed with the numerical values obtained with momentum space exact diagonalization. Right plot: Difference $\mathcal{C}_7(m+7)-\mathcal{C}_7(m)$ and the associated numerical values. The inset shows the conjectured values $\mathcal{C}_7(m)$ computed with two values of $a_0$ (respectively, $7.720$ and $7.716$) compatible with our estimate $a_0=7.718(2)$. In both the plots, $f(m)$ generically denotes one of the possible functions reported in the legend.}
    \label{math_conjecture_plots}
\end{figure}

To begin with, we plot the deviation of $\tilde{n}'_c$ from the line $y(m) \equiv 8m$, which is the integer upper bound of the linear regression extrapolated in Eq. \eqref{linear_fit_parameters_tildes}. We observe that the numerical data are organized in blocks of seven points, as highlighted in the left plot of Fig. \ref{math_conjecture_plots}, that can be characterized through an integer valued function. The characterization in terms of this function can be done for a specific sub-sequence of $m$ values or by considering all of them. We discuss both cases, showing how to conjecture the critical flux and highlighting similarities and differences.

Let us consider the first conjecture, which is for all values of $m$. Given the round-half up function, we notice that the estimate of $\tilde{n}'_c$ satisfies the inequalities
\begin{equation}
    \lfloor ma_0\rfloor+\bigg\lfloor a_1+\frac{1}{2}\bigg\rfloor\leq \tilde{n}'_c \leq \lfloor ma_0\rfloor+\bigg\lfloor a_1+\frac{1}{2}\bigg\rfloor+1\,.
\end{equation}
We further observe that $\lfloor a_1+1/2\rfloor=0$, due to our numerical estimates in Eq. \eqref{linear_fit_parameters_tildes}. Hence, motivated by the numerical results presented in Section \ref{critical_flux_estimation}, we conjecture that
\begin{equation}
    a_1+\frac{1}{2} = 0\,,
    \label{first_conjecture}
\end{equation} 
which implies $\tilde{n}'_c=\lfloor ma_0\rfloor$. Now, we can apply the Hermite identity \cite{Matsuoka1964}, a mathematical identity stating that, for any $x\in\mathbb{R}$ and $n\in\mathbb{N}$, it holds
\begin{equation}
    \lfloor xn\rfloor=\sum_{k=0}^{n-1}\bigg\lfloor x+\frac{k}{n}\bigg\rfloor\,.
    \label{hermite_identity}
\end{equation}
When applied to $\tilde{n}'_c(m)=\lfloor a_0m\rfloor$, the Hermite identity allows us to rewrite a generic element of the critical sequence in terms of a summation involving floor functions of the numerical parameter $a_0\in\mathbb{R}$, i.e. $\lfloor a_0 + k/m\rfloor$. Since $\lfloor a_0\rfloor=7$, the summation of Eq \eqref{hermite_identity} contains only the integers $\ell=7,8$. This is because the term $k/m$ in the argument of the floor function is limited in the range $[1/m,1)$, and for $m\geq1$ it can change the integer part of $a_0$ at most by one.

Due to this property, we can always identify a unique index, labelled as $\mathcal{C}_7$, such that $m-\mathcal{C}_7\leq m\{a_0\}\leq m-\mathcal{C}_7-1$, where $\{a_0\}\equiv a_0-\lfloor a_0\rfloor$ is the fractional part of $a_0$. We can then split the sum in the Hermite identity as
\begin{equation}
    \tilde{n}'_c=\sum_{k=0}^{m-1}\bigg\lfloor a_0+\frac{k}{m}\bigg\rfloor=\sum_{k=0}^{\mathcal{C}_7-1}\lfloor a_0\rfloor+\sum_{k=\mathcal{C}_7}^{m-1}(\lfloor a_0\rfloor+1)=8m-\mathcal{C}_7\,.
    \label{hermite_identity_tildenc_prime}
\end{equation}
The index $\mathcal{C}_7$ is an integer function of $m$, counting how many times the integer $\ell=7$ appears in the Hermite decomposition of $\lfloor a_0m\rfloor$.

With the introduction of this counting index $\mathcal{C}_7$, the critical flux $\Phi_c(m)$ for all values of $m$ can be written as
\begin{equation}
    \frac{\Phi_c(m)}{2\pi}=\frac{m}{8m-\mathcal{C}_7(m)}.
    \label{math_conjecture_phic}
\end{equation} 
Therefore, our conjecture \eqref{first_conjecture} implies that, given any integer $m\geq 1$ decomposed as a sum of the integers $\ell=7,8$, the associated critical flux $\Phi_c(m)$ is determined simply by counting how many times the number $\ell=7$ appear in such a decomposition, and then apply the Eq. \eqref{math_conjecture_phic}. 

It follows that, by the conjecture \eqref{first_conjecture}, the large-$m$ limit of $\Phi_c(m)$ is explicitly defined by
\begin{equation}
    \frac{\Phi_c}{2\pi}=\frac{1}{a_0}\,.
\end{equation}
However, since $a_0$ is not analytically known, it must be determined through fitting from the numerically determined values of $n(m)$, as done in Section \ref{critical_flux_estimation}. Thus, while the conjecture \eqref{first_conjecture} gives a form for $\Phi_c(m)$, it does not provide a new estimate for $\Phi_c$, which is instead provided by the second conjecture we provide below.

In Fig. \ref{math_conjecture_plots}, the left plot shows the value of the $\mathcal{C}_7(m)$ obtained by using \eqref{hermite_identity_tildenc_prime} with the numerical estimate for $a_0$ provided in \eqref{linear_fit_parameters_tildes}, and the numerically extracted one, i.e. the value $8m-\tilde{n}'_c$ for each $m$  (see Appendix \ref{table_critical_sequences}). From the plot, we also observe that the values of $\mathcal{C}_7(m)$ are grouped into blocks of length $7$. This feature can be captured by the quantity $\mathcal{C}_7(m+7)-\mathcal{C}_7(m)$, as depicted in the right plot of Fig. \ref{math_conjecture_plots}. 

An interesting observation is that for the sub-sequence of the form $m=m_p=7p+1$, with $p$ integer, the value of $\mathcal{C}_7(m_p)$, for $m\leq60$ (i.e. $p \leq8$), is given by $\mathcal{C}_7(m_p)=2p+1$. Furthermore, for these values $\mathcal{C}_7(m_{p+1})-\mathcal{C}_7(m_p)=2$. On the other hand, we point out that for all $m$ there may be isolated exceptions, as can be seen from the specific cases of $m=32$ ($\mathcal{C}_7$) and $m=39$ ($8m-\tilde{n}'_c$), and the clustering of the conjectured values depends on the estimate of $a_0$, which can modify the position of the spikes (see the inset in Fig. \ref{math_conjecture_plots}). This is a direct consequence of the fact that, despite its simplicity, our proposed conjecture in Eq. \eqref{first_conjecture} is valid up to the precision associated to $a_0$. Possible effects on the estimate of $\tilde{n}'_c$ are expected to show up for extremely large values of the integer $m$, without affecting however the convergence of the critical sequences to the critical value $\Phi_c$. The asymptotic relation of $\Phi_c$ to $a_0$ is given by Eq. \eqref{phi_c_a0_relation_asymptotic}. Lastly, we observe that the entire analysis can be applied to the other critical sequence $\tilde{n}_c$, with the working hypothesis $\lfloor a_1+1/2\rfloor=1$ within the precision of the parameter $a_1$ of Eq. \eqref{linear_fit_parameters_tildes}. 

We now state our second conjecture, which applies to a sub-sequence with integers $m$ of the form $m\;\text{mod}\;7=1$, such as $m=1,8,15,\ldots$. This sequence can be parametrized as $m_p=7p+1$, where $p\in\mathbb{N}$. For these values, we conjecture that
\begin{equation}
    \tilde{n}'_c-8m_p=2p+g(p),\qquad\qquad g(p):\quad \lim_{p\rightarrow\infty}\frac{g(p)}{2p}=0,
    \label{conjecture_subsequence}
\end{equation}
based on the numerical results obtained for $m\leq60$. For these specific values, i.e. when $p\leq 8$, the difference $\tilde{n}'_c-8m_p=2p+1$, and $g(p)$ is found to be equal to $1$. However, for larger values of $p$, deviations from the value $g(p)=1$, while still compatible with the conjecture, can be allowed. This can be exemplified already by a single case: consider $m_{19}=134$ with 
$p=19$, such that the associated $\tilde{n}_c'\times\tilde{n}'_c$ matrix is still computationally accessible, i.e. $\tilde{n}_c'=O(10^3)$, and the numerical considerations given at the end of Section \ref{energy_bands} apply. We find $\tilde{n}'_c=1034$ and $g(19)=2$. 

A complete characterization of $g(p)$ for very large values of $p$ is very challenging, due to increasingly higher matrix size in momentum space. Nevertheless, due to the co-prime condition on the pairs $(m,n)$, we conjecture that the function $g(p)$ is subdominant with respect to the linear term in $p$, leading to the assumption of Eq. \eqref{conjecture_subsequence}.

In light of these considerations and based on the conjecture \eqref{conjecture_subsequence}, the asymptotic critical flux can be estimated for large $p$, and it is given by
\begin{equation}
    \lim_{m_p\rightarrow\infty}\frac{m_p}{\tilde{n}'_c}=\lim_{p\rightarrow\infty}\frac{7p+1}{8(7p+1)-(2p+g(p))}=\frac{7}{54}\equiv \frac{\Phi_c^{{\rm (conj)}}}{2\pi}\,.    \label{subsequence_m_math_conjecture_phi_c}
\end{equation}
The value $7/54=0.1\overline{296}$ for $\Phi_c^{{\rm (conj)}}/2\pi$ is compatible with the value $\Phi_c/2\pi=0.1296(1)$ reported in Section \ref{critical_flux_estimation}.

\section{\label{conclusions}Conclusions}
We characterized the band structure of the three-dimensional (3D) Hofstadter model on cubic lattices in presence of isotropic fluxes, associated with a magnetic field ${\bf B}$ aligned along the main diagonal of the lattice and parametrized as a ratio of co-prime pairs $(m,n)$, i.e. $\Phi=2\pi \cdot m/n$. As a function of $m$, we identified the presence of a critical flux $\Phi_c(m)$, which separates two regimes characterized respectively by complete overlap between the degenerate bands of the model ($\Phi<\Phi_c$) and the presence of isolated band touching points between the $m$-th and $(m+1)$-th bands of the model ($\Phi>\Phi_c)$. 

By writing the model using the Hasegawa gauge in the magnetic Brillouin zone \cite{Hasegawa1990,BurrelloJPHYSMATH2017}, we performed numerical exact diagonalization and established a connection between the critical flux and the appearance of inversion points in the central bands of the model, along the $z$-axis of the reciprocal space. The minimum of the $(m+1)$-th band moves from $k_z=\pi$ to $k_z=0$ upon crossing the critical flux, passing from a band touching to an overlapping scenario. At the same time, we characterized the scaling of the ground state energy, which is a function of the magnetic flux $\Phi$. We showed that the Weyl energy, corresponding to the band touching points, does not scale with $m/n$ nor $m$. Its position remains, up to our numerical precision, the same for the explored critical pairs ($m \leq 60$).

Regarding the critical flux $\Phi_c$, we characterized it introducing four different critical sequences of integers $n$, as discussed in Section \ref{phic_definitions}. We distinguished two types of sequences: the first two, labelled as $n_c$ and $\tilde{n}_c$, tend towards $\Phi_c$ from below, i.e. from the overlapping scenario, while the remaining two, $n'_c$ and $\tilde{n}'_c$, approach it from above, in the bands touching case. All these sequences tends asymptotically to the same critical value, leading to the conclusion that $\Phi_c(m)$ is defined for any $m$ and that the limit for large $m$ is unique, well-defined and finite.

Considering the sequences $\tilde{n}_c$ and $\tilde{n}'_c$, we conjectured an analytical form for the critical flux as a function of the numerator $m$ parametrizing the magnetic field in two cases. If all the values of $m\in\mathbb{N}$ are considered, we applied the Hermite identity \cite{Matsuoka1964} to conjecture that $\tilde{n}_c'$ is determined by the difference $8m-\mathcal{C}_7(m)$, where $\mathcal{C}_7(m)$ is an index function counting the occurrences of the integer $\ell=7$ in the application of the Hermite identity. If, at variance, we consider only the subset of values for which $m\;\text{mod}\;7=1$, we provided a value for the asymptotic critical flux given by $\Phi_c^{{\rm (conj)}}/2\pi=7/54$. This prediction is based on the numerically determined values of $\tilde{n}'_c$ obtained through numerical ED for $m \leq 60$, but does not depend on the parameters estimated from the critical sequences.

%Experimental relevance
Knowing the existence and precise magnitude of the critical flux defined in this work for the 3D Hofstadter model is important to distinguish non-trivial topological regimes from trivial metallic ones \cite{Louvet2018,FontanaPRB2021} also in experimental scenarios. Indeed, the experimental realization of Hofstadter model with ultracold atoms with artificial gauge fluxes \cite{Aidelsburger2013,Miyake2013,Aidelsburger2015,Aidelsburger2018} combined with band-touching points detection techniques, such as interferometric experiments \cite{DucaScience2015}, Bragg spectroscopy \cite{GotzeNature2010} or Landau-Zener scattering processes \cite{LihKingPRL2012}, could be applied to verify the existence and separation of the overlapping and band-touching regimes. This can be further relevant for the identification of Weyl semimetallic phases and determination of Weyl nodes locations in momentum space \cite{CooperRMP2019}. It would be also useful to study the robustness of Weyl points when the hopping coefficients are not isotropic or for different orientation (parametrized by rational numbers) of the magnetic field.

%Mathematical importance
Finally, from the mathematical point of view, it would be very interesting in our opinion to complement our numerical observations and conjectures with analytical predictions of the critical flux $\Phi_c(m)$ based on the structure of the Jacobi matrix $\mathcal{H}({\bf k})$ in momentum space. This computation, which is not present in the literature to the best of our knowledge, can shed light on the existence and well-definiteness of the large-$m$ limit of $\Phi_c(m)$ and on the rational (or potentially irrational) nature of the large $m$ critical flux $\Phi_c$.

\section*{Acknowledgments}
We are very thankful to M. Burrello and M. Gallone for discussions and suggestions. P.F. gratefully acknowledges discussions with C. Iacovelli and M. Rizzi. We thank the Galileo Galilei Institute for Theoretical Physics for the hospitality and the INFN for partial support during the completion of this work. P.F. acknowledges the support of MCIN/AEI/10.13039/501100011033 (LIGAS PID2020-112687GB-C22), Generalitat de Catalunya (AGAUR 2021 SGR 00138) and the financial support of Unió Europea - NextGenerationEU.

\appendix

\section{\label{harper_eq}Harper equation}
Within the Hasegawa gauge \eqref{hasegawa_gauge}, the discrete Schr\"odinger equation $\mathcal{H}\psi(\bm{r})=E\psi(\bm{r})$ is written as 
\begin{align}
    \nonumber
    \frac{1}{2}\bigg[&\psi(\bm{r}+\hat{x})+\psi(\bm{r}-\hat{x})+\psi(\bm{r}+\hat{y})\exp{\bigg[i\bigg(x-y-\frac{1}{2}\bigg)\Phi\bigg]}+\psi(\bm{r}-\hat{y})\exp{\bigg[-i\bigg(x-y-\frac{1}{2}\bigg)\Phi\bigg]}\\
    & \psi(\bm{r}+\hat{z})\exp{[-i(x-y)\Phi]}+\psi(\bm{r}-\hat{z})\exp{[i(x-y)\Phi]}\bigg]=E\psi(\bm{r})\,,
\end{align}
where we used the Harper operator \cite{HarperPPS1955}
\begin{equation}
    H_{\hat{j}}\psi(\bm{r})=\frac{\psi(\bm{r}+a\hat{j})e^{i \theta_{j}\left({\bm r}\right)}+\psi(\bm{r}-a\hat{j})e^{-i \theta_{j}\left({\bm r}\right)}}{2}
\end{equation}
and set the lattice spacing $a=1$. The absence of the $z$-coordinates in the Peierls phases allows for the factorization of the wave function $\psi(\bm{r})=e^{ik_z z}\varphi(x,y)$, with $k_z\in[-\pi,\pi]$. The last two terms in the left-hand side of the discrete Schr\"odinger equation then become
\begin{equation}
    \psi(\bm{r}+\hat{z})\exp{[-i(x-y)\Phi]}+\psi(\bm{r}-\hat{z})\exp{[i(x-y)\Phi]}\qquad\rightarrow\qquad 2\cos{[k_z-\Phi(x-y)]}\varphi(\bm{r}')\,.
\end{equation}
With the flux defined in Eq. \eqref{rational_flux}, the Peierls phases $\theta_{x,y}(\bm{r})$ are periodic in $x,y$ with period $n$. We can apply the Bloch theorem to the wave function $\varphi(\bm{r}')$ \cite{Grosso_Parravicini_2000}, by writing
\begin{equation}
    \varphi(\bm{r}')=e^{i{\bf k}\cdot\bm{r}'}\tilde{\varphi}_{\ell{\bf k}}(\bm{r}'),\qquad{\bf k}\in\bigg[-\frac{\pi}{n},\frac{\pi}{n}\bigg]\times\bigg[-\frac{\pi}{n},\frac{\pi}{n}\bigg]\,.
\end{equation}
By inserting this into the Schr\"odinger equation we end up in the Harper equation for the 3D Hofstadter model
\begin{align}
    \nonumber
    \tilde{\varphi}_{\ell{\bf k}}(\bm{r}'+\hat{x})e^{ik_x}+\tilde{\varphi}_{\ell{\bf k}}(\bm{r}'-\hat{x})e^{-ik_x}&+\tilde{\varphi}_{\ell{\bf k}}(\bm{r}'+\hat{y})e^{i[k_y+\Phi(x-y-1/2)]}+\tilde{\varphi}_{\ell{\bf k}}(\bm{r}'-\hat{y})e^{-i[k_y+\Phi(x-y-1/2)]}\\
    &+2\tilde{\varphi}_{\ell{\bf k}}(\bm{r}')\cos{[k_z-\Phi(x-y)]}=E\tilde{\varphi}_{\ell{\bf k}}(\bm{r}')\,.
\end{align}
This eigenvalue equation has $n$ degenerate solutions, with ${\bf k}$ defined in the MBZ. As expected, we recover the same spectrum obtained with the momentum space diagonalization described in Section \ref{kspace_diagonalization_subsec}.

\section{\label{jacobi_inversion_points_App}Jacobi matrix for the band inversion points}
In this Appendix we report the periodic Jacobi matrix $\mathcal{H}({\bf k})$ for the values of ${\bf k}_{\text{min}}$ in Eqs. \eqref{jump_momentum_kmin_even_m}, \eqref{jump_momentum_kmin_odd_m}. When $m$ is even, we have $k_x=k_y=0$, simplifying the off-diagonal matrix elements to 
\begin{equation}
    U_j(0,0)\equiv U_j=1+e^{i(\Phi/2)}\varphi_j^*=1+e^{i\Phi(1/2-j)}\,,
\end{equation}
with the symmetry property $U_{n-j}(0,0)=1+e^{i\Phi(1/2-j)}e^{2\pi i m}=U_j(0,0)$.
The diagonal elements involve only $k_z$, which can be $k_z=0$ or $k_z=\pi$. In the two cases we have respectively
\begin{equation}
    D_j(0)\equiv D_j=2\cos(j\Phi),\qquad D_j(\pi)=2\cos(\pi-j\Phi)=-2\cos(j\Phi)=-D_j,
\end{equation}
with the symmetry property $D_{n-j}(0)=2\cos(2\pi m-j\Phi)=D_j(0)$. The matrix for $k_z=0$, which we label $\mathcal{H}_e$, has the simplified form
\begin{equation}
    \mathcal{H}_e=
    \begin{pmatrix}
        1 & 1+e^{-i\Phi/2} & 0 & \ldots & 0 & 1+e^{-i\Phi/2}\\
        1+e^{i\Phi/2} & 2\cos\Phi & 1+e^{-i3\Phi/2} & 0 & \ldots & 0\\
        0 & 1+e^{i3\Phi/2} & 2\cos2\Phi & 1+e^{-i5\Phi/2} & 0 & \ldots\\
        \vdots & \ddots & \ddots & \ddots & \ddots & \ddots \\
        \vdots & \ddots & \ddots & \ddots & \ddots & 1+e^{-i\Phi/2}\\
        1+e^{i\Phi/2} & 0 & \ldots & \ldots & 1+e^{i\Phi/2} & 2\cos\Phi\\
        \label{tridiagonal_matrix_even m}
    \end{pmatrix}\,;
\end{equation}
the corresponding one for $k_z=\pi$ has the elements on the diagonal with the opposite sign.

For odd values of $m$, the position of the minimum is at $(k_x,k_y)=(\bar{k},-\bar{k})$, therefore
\begin{equation}
    U_j(\bar{k},-\bar{k})=e^{-i\bar{k}}+e^{-i\bar{k}}e^{i\Phi(1/2-j)}=e^{-i\bar{k}}U_j\,,
\end{equation}
showing that the off-diagonal structure is the same as the even $m$ case, up to a phase factor due to the different location of the Weyl nodes in momentum space. The diagonal part, being $k_z=0,\pi$ for this case too, is the same as the even case. By calling $\mathcal{H}_o$ the matrix for odd $m$, we can write it for $k_z=0$ as
\begin{equation}
    \mathcal{H}_o=
    \begin{pmatrix}
        1 & e^{-i\bar{k}}(1+e^{-i\Phi/2}) & 0 & \ldots & 0 & e^{i\bar{k}}(1+e^{-i\Phi/2})\\
        e^{i\bar{k}}(1+e^{i\Phi/2}) & 2\cos\Phi & e^{-i\bar{k}}(1+e^{-i3\Phi/2}) & 0 & \ldots & 0\\
        0 & e^{i\bar{k}}(1+e^{i3\Phi/2}) & 2\cos2\Phi & e^{-i\bar{k}}(1+e^{-i5\Phi/2}) & 0 & \ldots\\
        \vdots & \ddots & \ddots & \ddots & \ddots & \ddots \\
        \vdots & \ddots & \ddots & \ddots & \ddots & e^{-i\bar{k}}(1+e^{-i\Phi/2})\\
        e^{-i\bar{k}}(1+e^{i\Phi/2}) & 0 & \ldots & \ldots & e^{i\bar{k}}(1+e^{i\Phi/2}) & 2\cos\Phi\\
        \label{tridiagonal_matrix_odd m}
    \end{pmatrix}\,,
\end{equation}
and similarly for $k_z=\pi$.

\section{\label{overlaps_analysis}Features of the critical sequences $n_c$ and $n'_c$}
\begin{figure}
    \centering
    \includegraphics[width=0.3\linewidth]{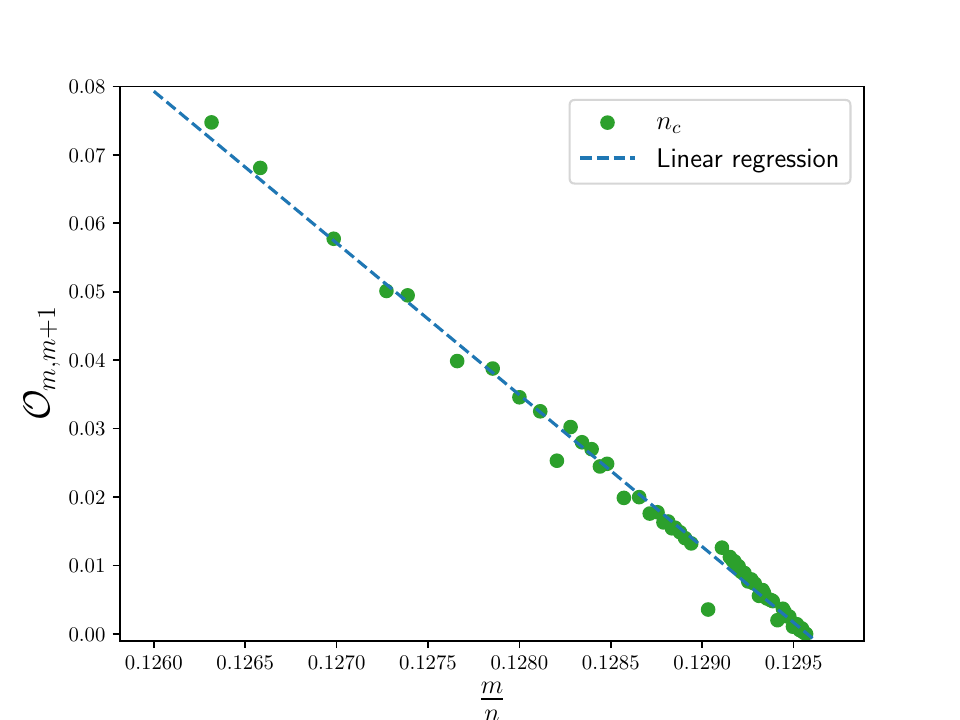}
    \qquad
    \includegraphics[width=0.28\linewidth]{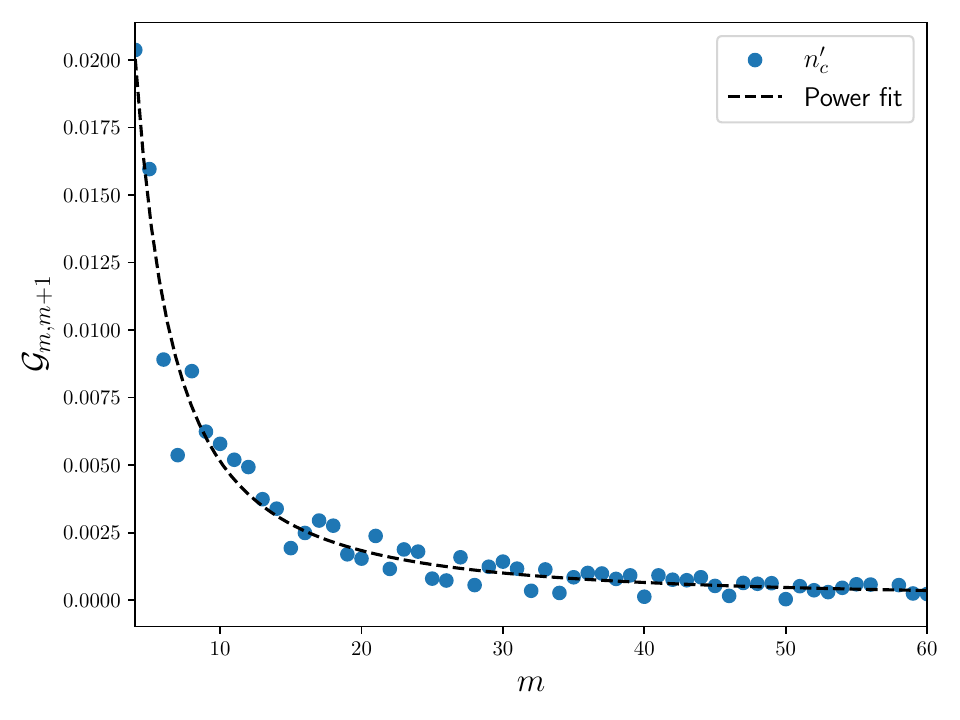}
    \qquad
    \includegraphics[width=0.33\linewidth]{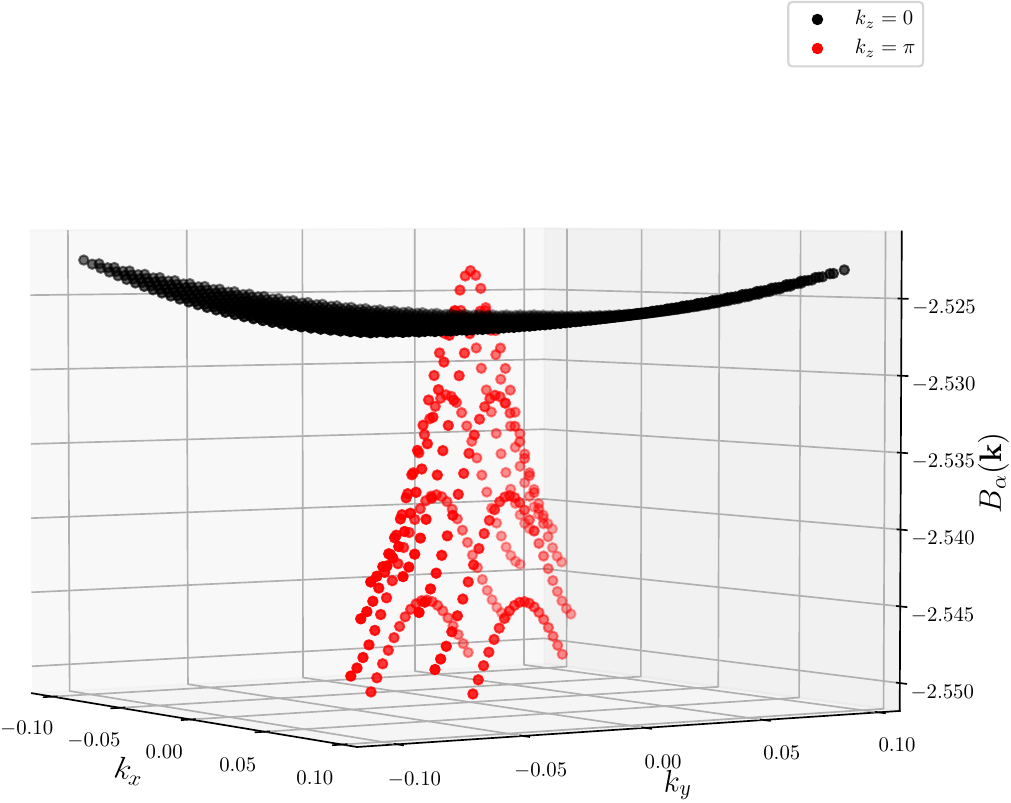}
    \caption{Left plot: Bands overlaps as a function of $m/n$ for the co-prime pairs $(m,n_c)$ listed in Table \ref{tab:critical_couples}, superimposed with the straight line $\mathcal{O}_{m,m+1}=m(\Phi/2\pi)+q$. Central plot: Bands gap as a function of $m/n$ for the co-prime pairs $(m,n'_c)$ listed in Table \ref{tab:critical_couples}, superimposed with the fit and parameters values of Eq. \eqref{gaps_powerfit}. Right plot: 3D scatter plot of overlapping bands, with $B_m(k_z=0)$ (red dots) and $B_{m+1}(k_z=\pi)$ (black dots), for $(m,n)=(4,31)$.}
    \label{overlap_linearreg_plot}
\end{figure}
If the critical flux is determined from below, i.e. $\Phi\nearrow\Phi_c$, it is convenient to look at the overlaps between the bands $B_{m}$ and $B_{m+1}$ \cite{Hasegawa1990}, choosing the first and third definitions of critical sequences ($n_c$, $\tilde{n}_c$) among the ones given at the end of Section \ref{phic_definitions}. Once the overlaps are identified, the corresponding integers in the other sequences ($n_c'$, $\tilde{n}_c'$) follow immediately. We repeat the same analysis done at the end of Section \ref{critical_flux_estimation}, this time considering the values of $n_c$ instead than $\tilde{n}_c$, that is only true co-prime pairs at fixed $m$. By performing the same linear regression of Eq. \eqref{linear_regression_overlaps}, we obtain the parameters $m=-22.3(3)$ and $q=2.88(4)$, with the associated estimate of the critical flux $\Phi_c/2\pi=m/n_c=0.129(1)$, consistent with the estimate $\Phi_c/2\pi\sim 4/31$ \cite{Hasegawa1990}. The corresponding bands overlap is showed in Fig. \ref{overlap_linearreg_plot} (right plot). We notice that this estimate is compatible with the one given in the main text, see Eq. \eqref{phi_c_estimates_twotilde_sequences}. 

If instead we consider $\Phi\searrow\Phi_c$, the bands of the model do not overlap, but touch in isolated points \cite{Hasegawa1990}. However, due to the numerical discretization of the explored MBZ and the consequent uncertainty on the energy values discussed in the manuscript, we observe small gaps separating the bands $B_m({\bf k})$, $B_{m+1}({\bf k})$. We denote such gaps by $\mathcal{G}_{m,m+1}=|\epsilon_{m+1}({\bf k})-E_m({\bf k})|$. Interestingly, we observe a regular behaviour of these gaps as a function of $m$, at a fixed value of the discretization $\delta k=O(10^{-3})$ in the %reduced 
MBZ. The plot of the data for the critical sequence $n'_c$ is showed in Fig. \ref{overlap_linearreg_plot} (central plot), superimposed with a fit of the form
\begin{equation}
    \mathcal{G}_{m,m+1}=\frac{g_0}{m^{g_1}},\qquad\qquad g_0=0.16(1),\quad g_1=1.49(9)\,.
    \label{gaps_powerfit}
\end{equation}

\section{\label{numerical_details_bands}Identification of the planes $\Pi_{\bar{k}_z}$}
\begin{figure}
    \centering
    \includegraphics[width=0.4\linewidth]{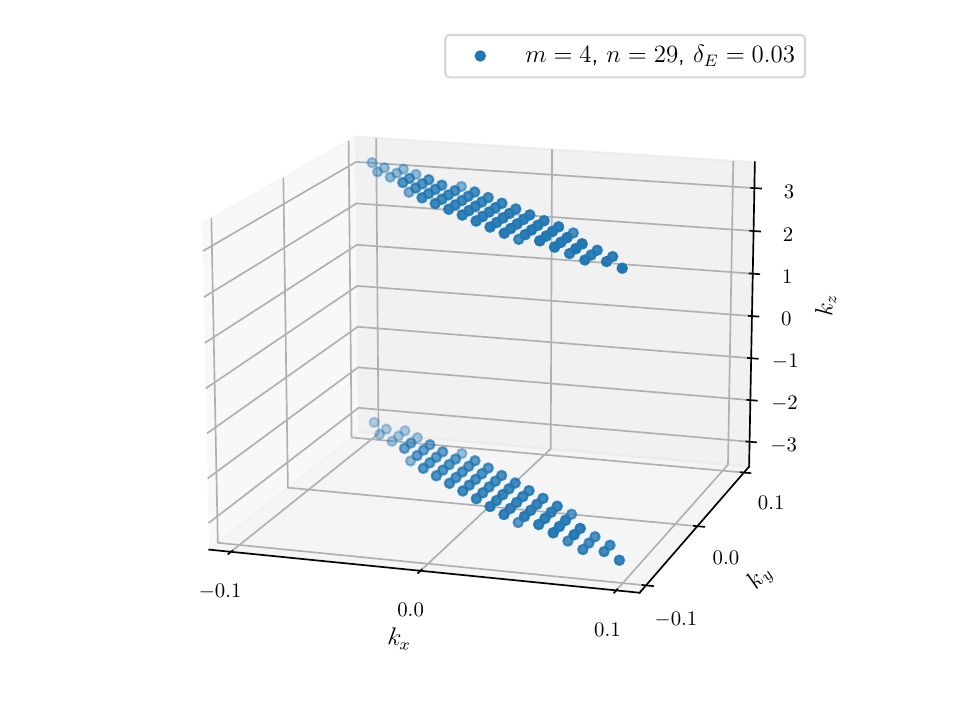}
    \qquad
    \includegraphics[width=0.4\linewidth]{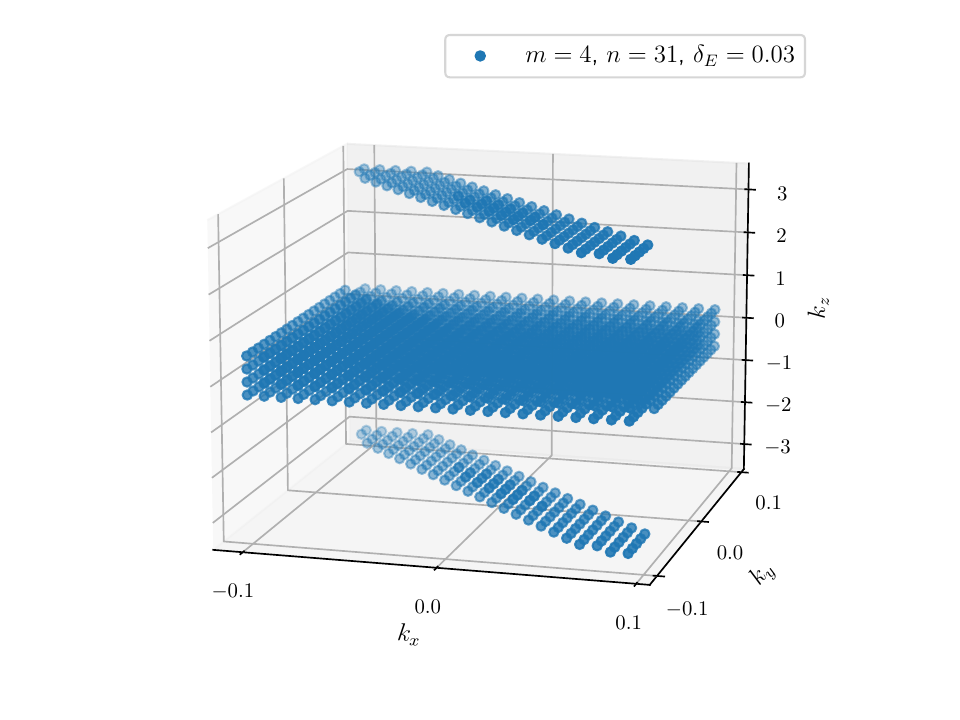}
    \caption{3D plot of the explored part of the MBZ in the intervals $I(\delta_E)$ around the band touching points for the co-prime pairs $(m,n)=(4,29)$ (left plot) and $(4,31)$ (right plot), with even $m$, for $\delta_E=0.03$. In both the plots, $\delta k=O(10^{-2})$.}
    \label{reducedMBZ_even m}
\end{figure}

\begin{figure}
    \centering
    \includegraphics[width=0.4\linewidth]{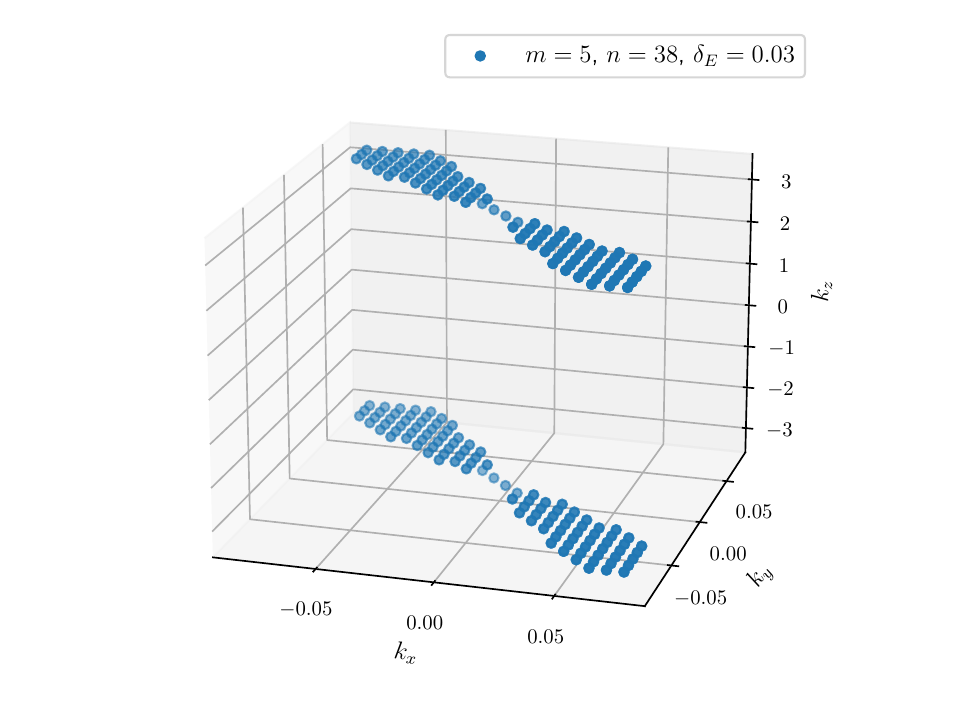}
    \qquad
    \includegraphics[width=0.4\linewidth]{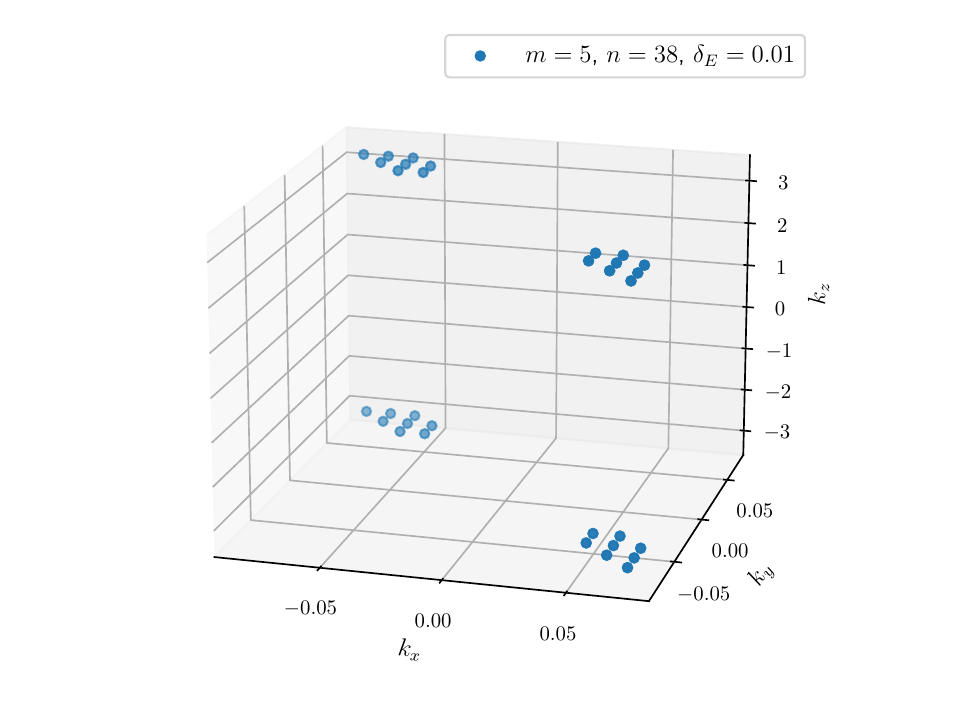}
    \caption{3D plot of the explored part of the MBZ in the intervals $I(\delta_E)$ around the band touching points for the co-prime pair $(m,n)=(5,38)$, with odd $m$, for two different values of $\delta_E=0.03$ (left plot), $\delta_E=0.01$ (right plot). In both the plots, $\delta k=O(10^{-2})$.}
    \label{reducedMBZ_odd m}
\end{figure}

In this Appendix we show the numerical details regarding the determination of the planes $\Pi_{\bar{k}_z}$ of Eq. \eqref{planes_reducedMBZ} in the main text. In the reduced ED procedure, we choose a discretization $\delta k$ of the MBZ and keep track of the momenta associated to $\epsilon_0=\epsilon_m$ and of the energy interval $I(\delta_E)=[E_m-\delta_E,\epsilon_{m+1}+\delta_E]$, where $\delta_E$ is a small parameter introduced to take into account the possible discretization effects related to $\delta k$. We plot the explored regions for even and odd $m$ in Figs.  \ref{reducedMBZ_even m} and \ref{reducedMBZ_odd m}, respectively, for different values of the parameter $\delta_E$. It is evident that by properly tuning $\delta k$ and $\delta_E$ the regions we are interested in are the bisectors $k_x=-k_y$ at $k_z=\pm\pi$ for odd $m$, with the additional plane at $k_z=0$ for even $m$ when the bands overlap (see the right plot of Fig. \ref{reducedMBZ_even m}). 
If we put in relation $\delta k$ to the linear size $L$ of the system in real space, and consider uniform grids in all the three directions in the MBZ, we have the usual correspondence \cite{Grosso_Parravicini_2000,Hasegawa1990}
\begin{equation}
    \delta k_{x,y}=\frac{2\pi}{nL},\qquad \delta k_z=\frac{2\pi}{L}\,.
\end{equation}
In our numerical diagonalization, we considered $\delta k_{x,y}=O(10^{-3})$, which for values of $n\in[8,500]$ is associated to linear sizes of the order $L\in[10,700]$. 

\section{\label{EWeyl_determination}Determination of the Weyl energy}
\begin{figure}
    \centering
    \includegraphics[width=0.4\linewidth]{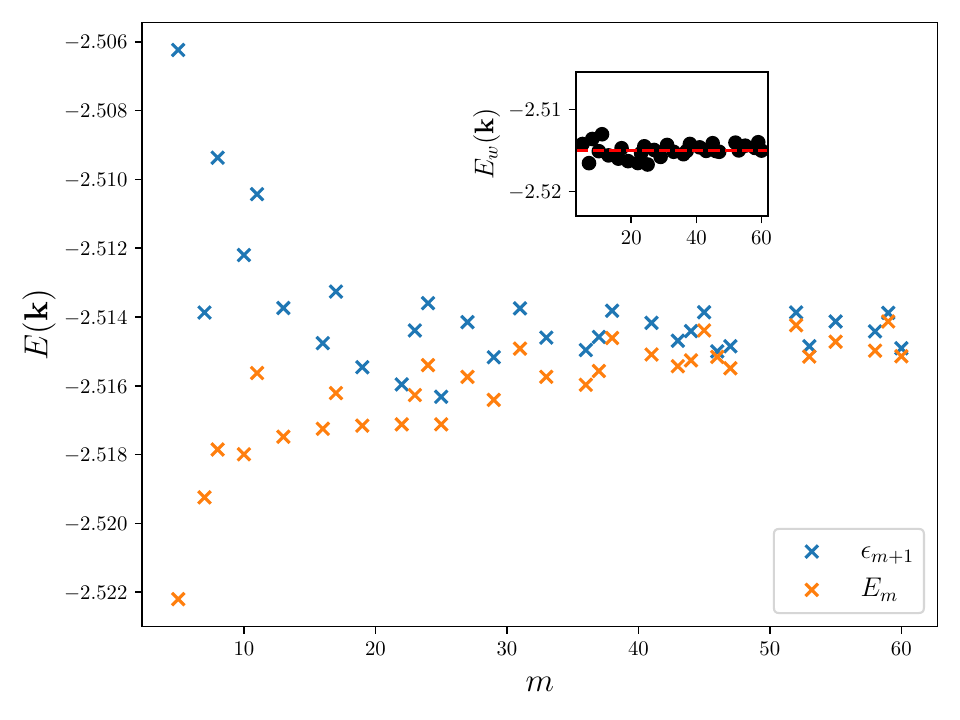}
    \qquad
    \includegraphics[width=0.4\linewidth]{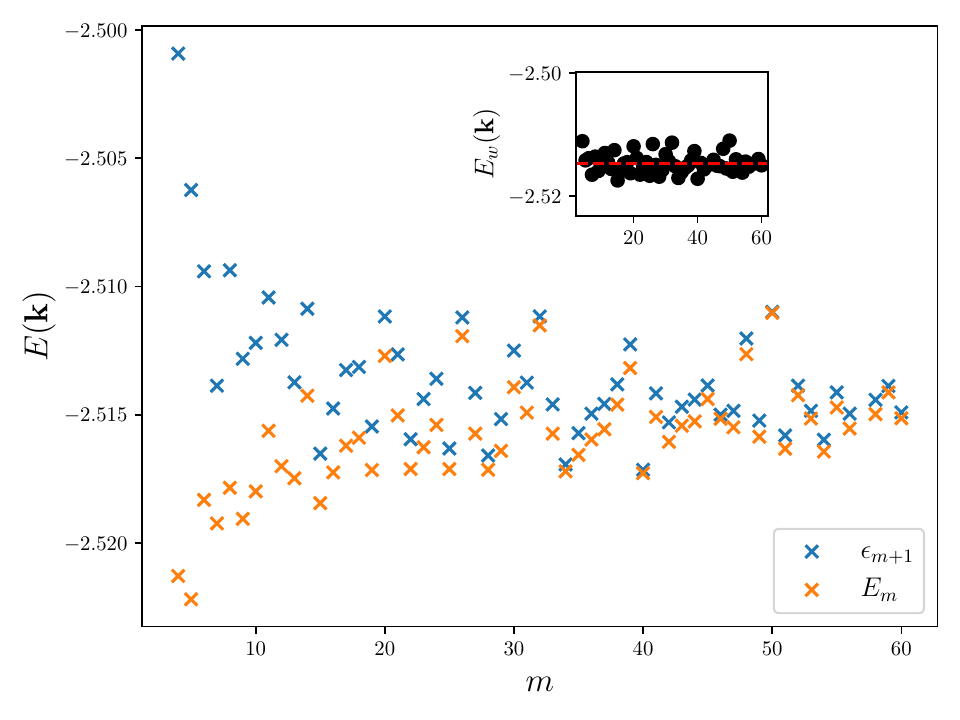}
    \caption{Energies $\epsilon_{m+1}({\bf k})$ and $E_m({\bf k})$ vs $m$ for the sequences $\tilde{n}_c'$ (left plot) and $n_c'$ (right plot). In both the plots, the insets show the estimated Weyl energies $E_w$ as a function of $m$ (black dots), with the corresponding estimates (red dashed lines).}
    \label{Ew_nc'_tildenc'_plots}
\end{figure}
We present here details about the estimate of $E_w({\bf k})$ from the critical sequence of integers $\tilde{n}'_c$. Due to the discretization effects discussed in Appendix \ref{overlaps_analysis}, we can identify the Weyl points $E_{w,m}$ for every co-prime pair $(m,\tilde{n}'_c)$ as the center of $\mathcal{G}_{m,m+1}$, i.e.
\begin{equation}
    E_{w,m}({\bf k})=\frac{\epsilon_{m+1}({\bf k})+E_m({\bf k})}{2}\,.
    \label{Ew_definition}
\end{equation}
We plot in Fig. \ref{Ew_nc'_tildenc'_plots} (left plot) the values of $\epsilon_{m+1}({\bf k})$ and $E_m({\bf k})$ as a function of $m$, highlighting the fact that the amplitude of $\mathcal{G}_{m,m+1}$ shrinks as long as we go progressively to large $m$. The asymptotic value of $E_{w,\infty}\equiv E_w$ extracted from this critical sequence can be estimated as the weighted average of all the values computed through Eq. \eqref{Ew_definition}, obtaining $E_w=-2.5150(9)$.

For comparison, we consider the critical sequence $n_c'$, whose energy values determining $\mathcal{G}_{m,m+1}$ are reported in Fig. \ref{Ew_nc'_tildenc'_plots} (right plot). In this case we observe the same behaviour, but with a larger standard deviation associated to the asymptotic Weyl energy, which turns out to be $E_w=-2.515(2)$, in any case compatible with the one obtained with the $\tilde{n}'_c$ sequence.

\newpage 

\section{\label{table_critical_sequences}Tables with sequences of integers $n(m)$}
\begin{table}[h!]
    \centering
    \begin{filecontents*}{nc_nc__tildenc_tildenc__couples_22dic.txt}
m   nc  nc' tildenc tildenc'
1	8	7		/		/
2	17	15		16		/
3	25	23		24		/
4	31	29		/		30
5	39	38		/		/
6	47	43		/		46
7	55	54		/		/
8	63	61		62		/	
9	70	68		/		69
10	79	77		78 		/
11	85	84		/		/
12	95	91		93 		92
13	101	100		/		/
14	109	107		/ 		108
15	116	113		/		115
16	125	123		124		/
17	132	131		/		/
18	139	137		/		138
19	147	146		/		/
20	157	153		155 	154
21	163	160		/		162
22	171	169		170		/
23	178	177		/		/
24	187	185		186		/
25	193	192		/		/	
26	201	199		/		200
27	209	208		/		/
28	219	215		217		216
29	224	223		/		/
30	233	229		232		231
\end{filecontents*}
\pgfplotstabletypeset[
        columns/m/.style={column name={$m$}},
        columns/nc/.style={
            column name=$n_c$,
        },
        columns/nc'/.style={
            column name=$n_c'$,
        },
        columns/tildenc/.style={
            column name={$\tilde{n}_c$},
        },
        columns/tildenc'/.style={
            column name={$\tilde{n}'_c$}
        },
        string type,
        before row=\hline,
        every last row/.style={after row=\hline},
        column type/.add={|}{},
        every last column/.style={column type/.add={}{|}},
    ]{nc_nc__tildenc_tildenc__couples_22dic.txt}
    \hspace{2cm}
    \begin{filecontents*}{nc_nc__tildenc_tildenc__couples_27dic.txt}
m   nc  nc' tildenc tildenc'
31	240	239		/		/
32	247	245		/		246
33	256	254		255		/
34	263	261		/		262
35	271	269		/		270
36	281	277		278		/
37	286	285		/		/
38	295	293		294		/
39	301	298		/		300
40	309	307		/		308
41	317	316		/		/
42	325	323		/		324
43	332	331		/		/
44	343	339		340		/
45	349	347		348		/
46	357	355		356		/
47	363	362		/		/
48	371	367		/		370
49	379	377		/		378
50	387	383		386		385
51	394	392		/		393
52	405	401		402		/
53	410	409		/		/
54	419	415		417		416
55	426	424		425		/
56	433	431		/		432
57	440	439		/		/
58	449	447		448		/
59	456	455		/		/
60	467	463		464		/
\end{filecontents*}
\pgfplotstabletypeset[
        columns/m/.style={column name={$m$}},
        columns/nc/.style={
            column name=$n_c$,
        },
        columns/nc'/.style={
            column name=$n_c'$,
        },
        columns/tildenc/.style={
            column name={$\tilde{n}_c$},
        },
        columns/tildenc'/.style={
            column name={$\tilde{n}'_c$}
        },
        string type,
        before row=\hline,
        every last row/.style={after row=\hline},
        column type/.add={|}{},
        every last column/.style={column type/.add={}{|}},
    ]{nc_nc__tildenc_tildenc__couples_27dic.txt}
    \caption{Critical pairs $(m,n)$, when defined, for the definitions $\{n_c,n'_c,\tilde{n}_c,\tilde{n}'_c\}$. The left table is for $1\leq m\leq30$, the right table for $31\leq m\leq60$.}
    \label{tab:critical_couples}
\end{table}

\newpage

\section{\label{table_energies_234digits}Tables of the rounded energies $E_m$, $\epsilon_{m+1}$}
\begin{table}[h!]
    \centering
    \begin{filecontents*}{m_tilde_ncprime_energies_2digits.txt}
m   nc  maxE1    minEm1
1	7	-2.57	-2.39
2	15	-2.54	-2.48
3	23	-2.53	-2.50
5	38	-2.52	-2.51
7	54	-2.52	-2.51
8	61	-2.52	-2.51
10	77	-2.52	-2.51
11	84	-2.52	-2.51
13	100	-2.52	-2.51
16	123	-2.52	-2.51
17	131	-2.52	-2.51
19	146	-2.52	-2.52
22	169	-2.52	-2.52
23	177	-2.52	-2.51
24	185	-2.52	-2.51
25	193	-2.52	-2.52
27	208	-2.52	-2.51
29	223	-2.52	-2.52
31	239	-2.51	-2.51
33	254	-2.52	-2.51
36	277	-2.52	-2.51
37	285	-2.52	-2.51
38	293	-2.51	-2.51
41	316	-2.52	-2.51
43	331	-2.52	-2.51
44	339	-2.52	-2.51
45	347	-2.51	-2.51
46	355	-2.52	-2.52
47	362	-2.52	-2.51
52	401	-2.51	-2.51
53	409	-2.52	-2.51
55	424	-2.51	-2.51
57	439	-2.52	-2.52
58	447	-2.51	-2.51
59	455	-2.51	-2.51
60	463	-2.52	-2.51
\end{filecontents*}
\pgfplotstabletypeset[
        columns/m/.style={column name={$m$}},
        columns/nc/.style={
            column name=$\tilde{n}'_c$,
        },
        columns/maxE1/.style={
            column name=$E_m$,
        },
        columns/minEm1/.style={
            column name={$\epsilon_{m+1}$},
        },
        string type,
        before row=\hline,
        every last row/.style={after row=\hline},
        column type/.add={|}{},
        every last column/.style={column type/.add={}{|}},
    ]{m_tilde_ncprime_energies_2digits.txt}
    \hspace{2.5cm}
    \begin{filecontents*}{m_tilde_ncprime_energies_3digits.txt}
m   nc  maxE1    minEm1
1	7	-2.567	-2.392
2	15	-2.539	-2.482
3	23	-2.529	-2.498
5	38	-2.522	-2.506
7	54	-2.519	-2.514
8	61	-2.518	-2.509
10	77	-2.518	-2.512
11	84	-2.516	-2.510
13	100	-2.517	-2.514
16	123	-2.517	-2.515
17	131	-2.516	-2.513
19	146	-2.517	-2.515
22	169	-2.517	-2.516
23	177	-2.516	-2.514
24	185	-2.515	-2.514
25	192	-2.517	-2.516
27	208	-2.516	-2.514
29	223	-2.516	-2.515
31	239	-2.515	-2.514
33	254	-2.516	-2.515
36	277	-2.516	-2.515
37	285	-2.516	-2.515
38	293	-2.515	-2.514
41	316	-2.515	-2.514
43	331	-2.515	-2.515
44	339	-2.515	-2.514
45	347	-2.514	-2.514
46	355	-2.515	-2.515
47	362	-2.515	-2.515
52	401	-2.514	-2.514
53	409	-2.515	-2.515
55	424	-2.515	-2.514
57	439	-2.515	-2.515
58	447	-2.515	-2.514
59	455	-2.514	-2.514
60	463	-2.515	-2.515
\end{filecontents*}
\pgfplotstabletypeset[
        columns/m/.style={column name={$m$}},
        columns/nc/.style={
            column name=$\tilde{n}'_c$,
        },
        columns/maxE1/.style={
            column name=$E_m$,
        },
        columns/minEm1/.style={
            column name={$\epsilon_{m+1}$},
        },
        string type,
        before row=\hline,
        every last row/.style={after row=\hline},
        column type/.add={|}{},
        every last column/.style={column type/.add={}{|}},
    ]{m_tilde_ncprime_energies_3digits.txt}
    \hspace{2.5cm}
    \begin{filecontents*}{m_tilde_ncprime_energies_4digits.txt}
m   nc  maxE1    minEm1
1	7	-2.5670	-2.3923
2	15	-2.5393	-2.4822
3	23	-2.5291	-2.4983
5	38	-2.5222	-2.5062
7	54	-2.5192	-2.5139
8	61	-2.5178	-2.5094
10	77	-2.5180	-2.5122
11	84	-2.5156	-2.5104
13	100	-2.5175	-2.5137
16	123	-2.5173	-2.5148
17	131	-2.5162	-2.5133
19	146	-2.5172	-2.5155
22	169	-2.5171	-2.5160
23	177	-2.5163	-2.5144
24	185	-2.5154	-2.5136
25	192	-2.5171	-2.5163
27	208	-2.5157	-2.5141
29	223	-2.5164	-2.5152
31	239	-2.5149	-2.5138
33	254	-2.5157	-2.5146
36	277	-2.5160	-2.5150
37	285	-2.5156	-2.5146
38	293	-2.5146	-2.5138
41	316	-2.5151	-2.5142
43	331	-2.5154	-2.5147
44	339	-2.5153	-2.5144
45	347	-2.5144	-2.5139
46	355	-2.5152	-2.5150
47	362	-2.5155	-2.5148
52	401	-2.5142	-2.5139
53	409	-2.5151	-2.5149
55	424	-2.5147	-2.5141
57	439	-2.5154	-2.5150
58	447	-2.5150	-2.5144
59	455	-2.5141	-2.5139
60	463	-2.5151	-2.5149
\end{filecontents*}
\pgfplotstabletypeset[
        columns/m/.style={column name={$m$}},
        columns/nc/.style={
            column name=$\tilde{n}'_c$,
        },
        columns/maxE1/.style={
            column name=$E_m$,
        },
        columns/minEm1/.style={
            column name={$\epsilon_{m+1}$},
        },
        string type,
        before row=\hline,
        every last row/.style={after row=\hline},
        column type/.add={|}{},
        every last column/.style={column type/.add={}{|}},
    ]{m_tilde_ncprime_energies_4digits.txt}
    \caption{Maximum of the $m$-th band ($E_m$) and minimum of the $(m+1)$-th band ($\epsilon_{m+1}$) for the sequence $\tilde{n}'_c$. From left to right: Energies rounded to the second, third and fourth significant digits, respectively.}
    \label{tab:energies_234digits}
\end{table}

\bibliography{biblio}

\end{document}